# Tuneable electronic properties in graphene

M.F.Craciun[1*†], S.Russo[2*], M.Yamamoto[3] and S.Tarucha[3]

[1]*Centre for Graphene Science, School of Engineering, University of Exeter, Exeter, UK*
[2]*Centre for Graphene Science, School of Physics, University of Exeter, Exeter, UK*
[3]*Department of Applied Physics, University of Tokyo, Tokyo, Japan*

**Abstract**
Novel materials are in great demand for future applications. The discovery of graphene, a one atom thick carbon layer, holds the promise for unique device architectures and functionalities exploiting unprecedented physical phenomena. The ability to embed graphene materials in a double gated structure allowed on-chip realization of relativistic tunneling experiments in single layer graphene, the discovery of a gate tunable band gap in bilayer graphene and of a gate tunable band overlap in trilayer graphene. Here we discuss recent advances in the physics and nanotechnology fabrication of double gated single- and few-layer graphene devices.

**Introduction**

Graphene – a monolayer of carbon atoms arranged in a honeycomb lattice – is a novel material with an unprecedented combination of physical properties [1-7]. For instance, graphene is transparent [8], conducting [3, 4], bendable [9, 10] and yet it is one of the strongest known materials [11]. It is therefore not surprising that graphene-based electronics holds great promise for future applications. Apart from the high technological potential, the relativistic nature of charge carriers in graphene provides the opportunity to address fundamental questions in condensed matter physics not accessible in any other material [12-17]. Unlike massive charge carriers governing conduction in common semiconductor materials, electrons and holes in graphene obey a linear energy dispersion relation (see Fig. 1a) and behave as chiral massless particles: i.e., Dirac fermions [6,18]. These unusual properties lie at the origin of a number of novel physical phenomena, such as an unconventional quantization sequence in the quantum Hall regime [3, 4], Klein tunneling [19-23] and Veselago lensing [24].

The already rich variety of physical phenomena accessible in monolayer graphene becomes yet more valuable when considering graphene as part of a larger family of materials: Few Layer Graphene (FLG). Indeed, recent experiments revealed that bilayer graphene (Fig.1b) is the only known material system to exhibit a gate-tunable band gap [25-35], whereas trilayer graphene (Fig.1c) is the only known semimetal with a gate tunable overlap between the conduction and the valence bands [36]. These discoveries demonstrate that each member of the FLG family is a unique material system with its own potential for device applications.

Recent advances in double-gated transistor architectures demonstrated that this device configuration is a highly flexible platform for investigating the electronic properties of FLG. In these structures, graphene-materials are sandwiched between two electrostatic gates which have a dual

---

[*] Both authors contributed equally to this work
[†] Corresponding author: M.F.Craciun@exeter.ac.uk; address: Harrison building North Park Road, Exeter EX4 4QF, (UK); tel: +44 1392 724049

valency, i.e. they are used to independently control the local charge type and density of graphene and to impose a perpendicular electric field onto the FLG rendering energetically inequivalent its layers. Respectively, the ability to control locally the charge type allowed the creation of p-n junctions with tunable junction polarity in monolayer graphene [21-23, 37-42]. Whereas, the capability to impose a perpendicular electric field onto FLG granted access to the electric field tunable band structure in bilayers [26-35] and trilayers [36]. Given the relevance and the large number of experiments investigating a variety of physical phenomena in these double gated graphene structures, it is timely to review the fundamental achievements in this research area.

In this review we will discuss the physics and nano-fabrication of double gated single- and few-layer graphene devices. First, we will introduce few basic electronic properties of graphene and FLG relevant for the understanding of the reviewed experiments. Then we will review the development of various fabrication technologies employed to realize double gated devices. We will proceed to summarize the novel physical phenomena accessible in double gated structures originated by the sole local control of charge type and density. Finally we will discuss the tunability of FLG low energy band dispersion by means of an external perpendicular electric field applied onto these material systems.

**Electronic properties of graphene materials**

The low energy band dispersion of graphene can be calculated using a tight binding model for electrons hopping in the honeycomb lattice [18]. The Bravais lattice of graphene consists of a unit cell with a basis of two carbon atoms (*A* and *B* in Fig. 1a). In a first approximation, we consider only hopping between nearest neighbour atomic sites since higher order hopping terms are significantly smaller. This results in two independent point per Brillouin zone, *K* and *K'* –also known as valleys, where the valence and conduction bands touch. The electronic states close to the Fermi level (*E=0*, see Fig. 1b) can be independently described by an effective Hamiltonian for each of these two high symmetry *K* and *K'* points. Respectively, $H_K$ and $H_{K'}$ are:

$$H_K = \hbar v_F \begin{pmatrix} 0 & k_x - ik_y \\ k_x + ik_y & 0 \end{pmatrix} \quad \text{(eq.1)}$$

$$H_{K'} = -\hbar v_F \begin{pmatrix} 0 & k_x + ik_y \\ k_x - ik_y & 0 \end{pmatrix} \quad \text{(eq.2)}$$

where both Hamiltonians operate on $\varphi = (\varphi_A, \varphi_B)^T$, with $\varphi_A$ and $\varphi_B$ the electron/hole Bloch wave functions components on the *A* and *B* sublattices [43]. This two-component wave function resembles the spinor wavefunctions in quantum electrodynamics, where the index of the spin corresponds to the sublattice for graphene and is referred to as pseudospin. The resulting low energy band dispersion in graphene is linear $E_k = \pm \hbar v_F |k|$, revealing that charge particles in graphene are mass-less and move with an energy independent Fermi velocity of $10^6$ m/s [3, 4, 44,45] (Fig. 1b).

In each valley the particles wave function is an eigen function of the helicity operator, meaning that in the *K* point the pseudospin direction is parallel to the momentum for electrons (states with $E_k > 0$) and antiparallel for holes (states with $E_k < 0$). Whereas, at the *K'* point the pseudospin direction is antiparallel to the momentum for electrons and parallel for holes. Therefore, if we consider charge transport through structures where the pseudospin is conserved –i.e. in the absence of short range disorder potential- the backscattering is strictly forbidden since it would require intervalley scattering from the *K* to *K'* point [46, 47]. This helicity (or chirality) of Dirac fermions in graphene plays a central role for the Klein tunneling effect as it will be shown later.

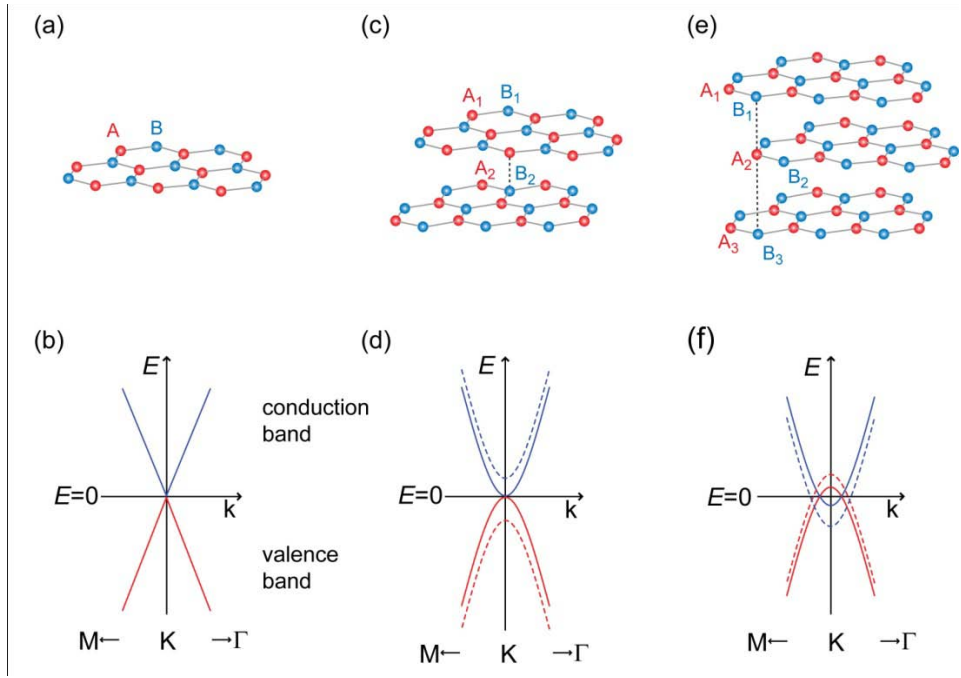

**Figure 1.** Tilted view of the crystal structure (top) and schematic band structure (bottom) of monolayer graphene ((a) and (b)), bilayer graphene ((c) and (d)) and trilayer graphene ((e) and (f)). The unit cell contains two equivalent carbon atoms –*A* and *B*– for monolayer graphene (a), four atoms –$A_1$, $B_1$, $A_2$ and $B_2$– for bilayer graphene (c) and six atoms –$A_1$, $B_1$, $A_2$, $B_2$, $A_3$ and $B_3$– for trilayer graphene (e). For Bernal (or *AB*) stacked bilayer the top layer has its $A_1$ atom on top of the $B_2$ atom of the bottom layer, see the dashed line in (c). Bernal (or *ABA*) stacked trilayer has the atoms in $B_1$, $A_2$ and $B_3$ aligned on top of each other, see the dashed line in (e). The continuous lines in (b), (d) and (f) represent the schematic band structure of graphene-materials in the absence of a perpendicular electric field, whereas the dotted lines show the resulting band structure upon application of a perpendicular electric field. Note that the perpendicular electric field has no effect on the band structure of monolayer graphene.

Though thicker few layer graphene material systems might be naively considered as the parallel of several single layer graphene, studies of the low energy electronic band structures of FLG demonstrate that each specific thickness of FLG is a unique material system remarkably different from a single layer. Here we will consider few-layer graphene with Bernal stacking -most often found in natural graphite- where A atoms in one layer are on top of B atoms of an adjacent layer, see Fig. 1.

The unit cell of bilayer *AB*-stacked graphene consists of a basis of four atoms labelled $A_1$, $B_1$, $A_2$ and $B_2$ belonging to different atomic planes as indicated by the numerical index (Fig. 1c). Most of the low energy band dispersion characteristics are properly described by a tight binding calculation where the in-plane nearest neighbour coupling (i.e. hopping from $A_1$ to $B_1$ and $A_2$ to $B_2$ atomic site) and the interlayer coupling between $A_2$ and $B_1$ atoms are considered [48]. The bilayer' band dispersion has a set of four parabolic bands in each of the two high symmetry *K* and *K'* points. If the onsite energy for an electron/hole particle is independent of the atomic site, in the low energy limit the effective Hamiltonians [48-51] for these bands become:

$$H_K = -\frac{\hbar^2}{2m}\begin{pmatrix} 0 & (k_x - ik_y)^2 \\ (k_x + ik_y)^2 & 0 \end{pmatrix} \quad \text{(eq.3)}$$

$$H_{K'} = -\frac{\hbar^2}{2m}\begin{pmatrix} 0 & (k_x + ik_y)^2 \\ (k_x - ik_y)^2 & 0 \end{pmatrix} \quad \text{(eq.4)}$$

which operate on the two Bloch wavefunction components $\varphi = (\varphi_{A1}, \varphi_{B2})^T$ on the sublattices $A_1$ and $B_2$ located in different layers and with $m$ the effective mass. The resulting two low energy parabolic bands of bilayer graphene touch each other at zero energy, making bilayer a two dimensional zero-gap semiconductor. Whenever the energetic equivalence between sublattices $A_1$ and $B_2$ is broken, the diagonal terms in the effective Hamiltonians of eq. 3 and eq. 4 are non zero and equal to the difference in the onsite energy between $A_1$ and $B_2$. The breaking of the interlayer symmetry in bilayer graphene pushes the highest valence and lowest conduction bands apart leading to the opening of a band-gap in the $K$ and $K'$ points as shown by the dashed lines in Fig. 1d [26,49, 52].

Bernal stacked trilayer graphene has the atoms in $B_1$, $A_2$ and $B_3$ aligned on top of each other (Fig. 1e). The trilayers low energy band dispersion is yet unique [36, 50, 53-61]. In particular, trilayer graphene is the thinnest of the few layer graphene material systems in which all the hopping parameters that play a role in the band structure of bulk graphite first appear. Recent tight binding calculations of trilayer graphene band structure have shown that the top of the valence band and bottom of conduction band overlap, leading to a finite density of states at the Fermi level [59, 62, 63]. When the interlayer symmetry of trilayer is broken, for instance by means of an external perpendicular electric field, the low energy parabolic electron and hole bands shift to lower energies. In a disorder-broadened energy window this leads to an increase in the expectation value of the band velocity at the Fermi level ($E=0$) [59, 62, 63], experimentally visible as an increase of energy overlap between conduction and valence band [33, 36] (see Fig. 1f). This is the opposite as the band gap opening induced by the interlayer symmetry breaking in bilayer graphene.

Theoretical calculations of the gate-tunable low energy band structure of Bernal stacked FLG with more than three layers [62- 64] predict that breaking the interlayer symmetry of these FLG opens an energy gap at the Fermi level, independently of the number of layers. However, for an even number of layers this gap is larger than the few meV gap predicted in the case of an odd number of graphene layers. To date these predictions have not been confirmed experimentally since FLG electronic properties are largely unexplored. Also experimentally unexplored are the electronic properties of FLG with different stacking than Bernal, such as the rhombohedral stacking [65]. In this case, surface-state bands are predicted to dominate the low energies causing a strong non-linear screening effect through the opening of an even larger energy gap than what expected for Bernal stacked FLG. Therefore, rhombohedral FLG are possibly better suited for transistor applications than Bernal stacked FLG.

Another fascinating property of graphene and its few layers is the possibility to continuously drive the Fermi level from the valence to the conduction band simply by applying a gate voltage, resulting in a pronounced ambipolar electric field effect [1, 2]. This is shown in Fig. 2 for graphene: as the Fermi level is driven inside the conduction (valence) band, the conductivity σ increases with increasing the concentrations of electrons (holes) induced by positive (negative) gate voltages. Whenever the Fermi level goes from the conduction (valence) band to the valence (conduction) band it crosses the zero density of states point – i.e. the Dirac point. Noticeably, since the early discovery of graphene, charge transport experiments in transistor structures revealed that even if the carrier density vanishes at the Dirac point, the conductivity does not go to zero but it remains finite at a value of

~$4e^2/h$ (Fig. 2) [3, 5, 67, 68]. Although a minimum conductivity at the neutrality point has been theoretically predicted for Dirac electrons in graphene in the ballistic regime with a value of $4e^2/\pi h$, the origin of the observed minimum conductivity and its interplay with disorder in diffusive devices is currently fuelling both theoretical and experimental debate [6, 66].

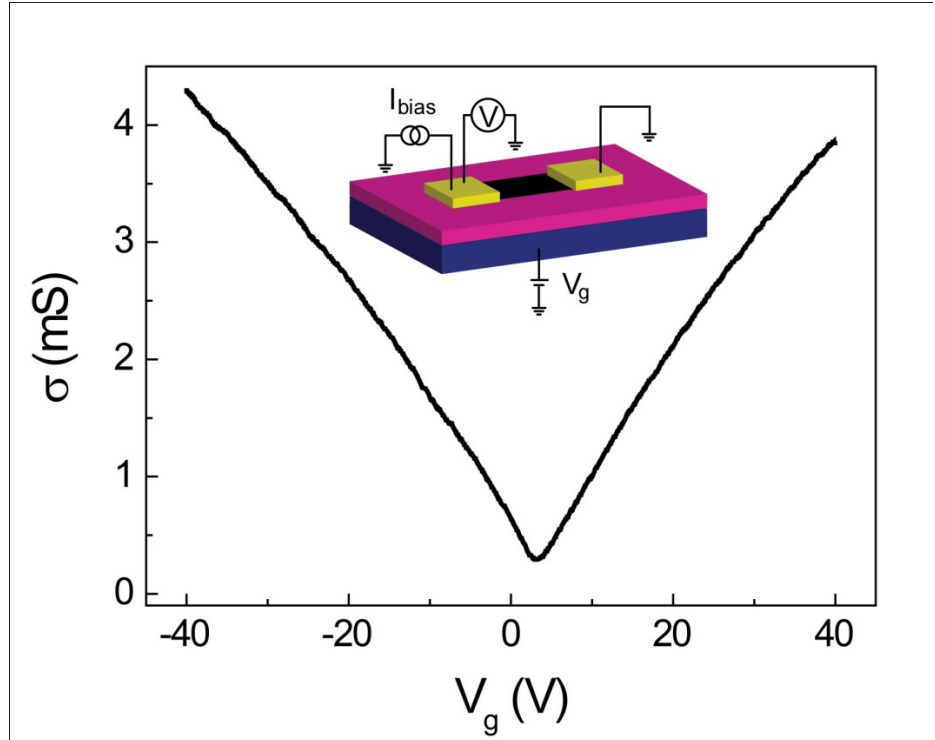

**Figure 2.** Ambipolar electric field effect in a graphene transistor. The inset shows a schematic view of the transistor structure. The metallic electrodes (yellow) on top of graphene (black) are the transistor's source and drain, and the Si substrate (dark blue) covered by $SiO_2$ (pink) acts as the gate. The position of the Fermi level in graphene is controlled by the voltage applied to the gate, $V_g$. The conductivity σ is determined by applying an ac-current bias $I_{bias}$ and measuring the resulting voltage V across the device. σ increases with increasing $V_g$ for both gate polarities, indicating that electrons (holes) are induced by positive (negative) gate voltages. As the Fermi level is driven through the Dirac point, σ remains finite even though the carrier density vanishes.

When Dirac fermions in graphene travel in a perpendicular magnetic field (*B*), they experience the Lorentz force which bends their trajectory. In the quantum regime, these cyclotron orbits give rise to discrete energy levels (Landau levels) which are remarkably different in the case of single-layer graphene as compared to conventional two-dimensional electron gases. Due to the unique chiral nature of particles in graphene, the Landau levels appear at energy values $E_N = \pm v_F\sqrt{2e\hbar BN}$, with *N* an integer number [69,70]. In graphene a shared Landau level between electrons and holes is present at *E=0*. These Landau levels energy values are at the origin of the observed quantization sequence of the Hall conductance, i.e. $G_{xy} = \pm(4e^2/h)(N + 1/2)$ where the factor 4 accounts for spin and valley degeneracy (Fig. 3a) [71, 72]. In bilayer graphene the Landau levels are yet different from single-layer graphene and also from common conductors. In bilayers $E_N = \pm\hbar\omega_c\sqrt{N(N+1)}$ with $\omega_c = eB/m$ and the quantization sequence of the Hall conductance is $G_{xy} = \pm 4Ne^2/h$ for $N \geq 1$ (Fig. 3b) [49]. The double quantization step at $E=0$ is a unique property of bilayer graphene, and together with the different values of the quantized conductance can be used to distinguish bilayers from single-layers.

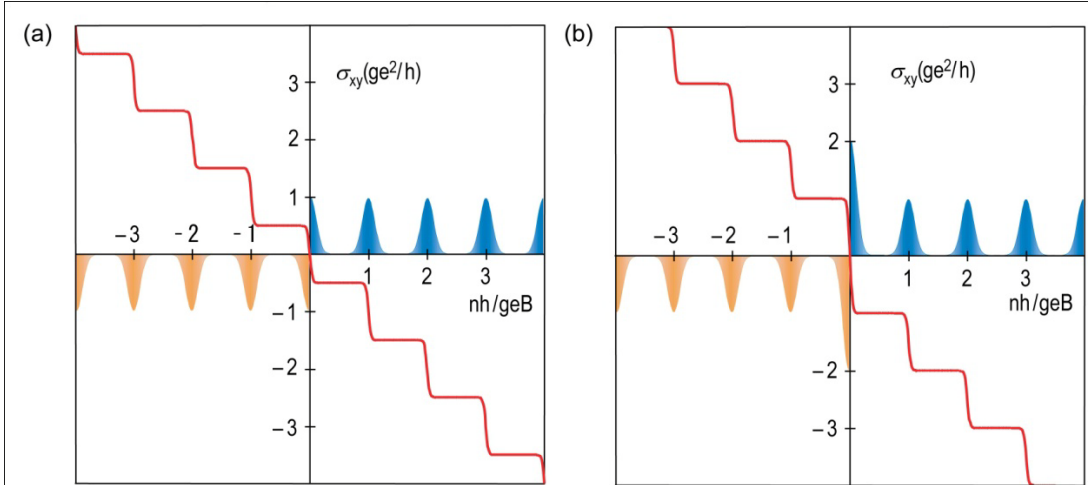

**Figure 3.** Quantum Hall Effect in monolayer graphene (a) and bilayer graphene (b), from [73]. Copyright 2006 Nature Publishing Group. Plateaus in Hall conductivity $\sigma_{xy}$ occur for monolayer at values $(ge^2/h)(N+1/2)$ with $N$ an integer whereas for bilayer at values $(ge^2/h)N$ with $N\geq 1$ an integer. Here $e^2/h$ is the conductance quantum and $g=4$ is the system degeneracy. The distance between steps along the concentration axis is defined by the density of states $geB/h$ on each Landau level with $B$ the magnetic field and $h/e$ the flux quantum. The corresponding sequences of Landau levels as a function of carrier concentrations $n$ are shown in blue and orange for electrons and holes, respectively.

The degeneracy of the Landau levels coincides with the number of magnetic flux quanta enclosed through the sample ($N_\phi = \frac{BL_xL_y}{\phi_0}$, with $L_x$ and $L_y$ dimensions of the samples and $\phi_0$ magnetic flux quanta). For a given value of the Fermi level we indicate the number of electrons by $N_{el}$, we can determine how many Landau levels are filled by defining the filling factor $\nu = \frac{N_{el}}{N_\phi}$. Note that in the integer quantum Hall effect the plateaus of conductance occur around integer values of $\nu$. In the bulk of the sample, clockwise and counterclockwise pieces of cyclotron orbits overlap and cancel, leading to a vanishing of current in the bulk. At the edges, the orbits are truncated in response to the confining potential created by the boundary and give rise to an edge current. This current arising from states at the edge is a hallmark of quantum Hall systems.

**Fabrication of double-gated graphene devices**

In all the experiments discussed in this review, graphene-based materials (e.g. monolayer, bilayer, etc..) are embedded in a sandwich structure between two electrostatic gates: a back gate that extends over the entire flake and a top gate that can also cover the whole flake as well as smaller predefined regions of the flake leading to a local-gating action, see Fig. 4a and b. This double-gated device architecture has the specific purpose to control simultaneously and independently the Fermi Energy (i.e., charge type and density) and the perpendicular electric field applied onto the graphene-materials. At first graphene is produced by mechanical exfoliation of bulk graphite and transferred onto 300nm thick $SiO_2$ layer thermally grown on a highly conductive Si substrate which acts as a back gate [2-4]. Whereas, top gates are subsequently fabricated by deposition onto the flake of an oxide dielectric layer (e.g. $SiO_2$, $Al_2O_3$, $HfO_2$, $Y_2O_3$ air, etc..) and of a metal layer acting as a top electrode [21, 22, 27, 30, 33-38, 40-42, 74-87].

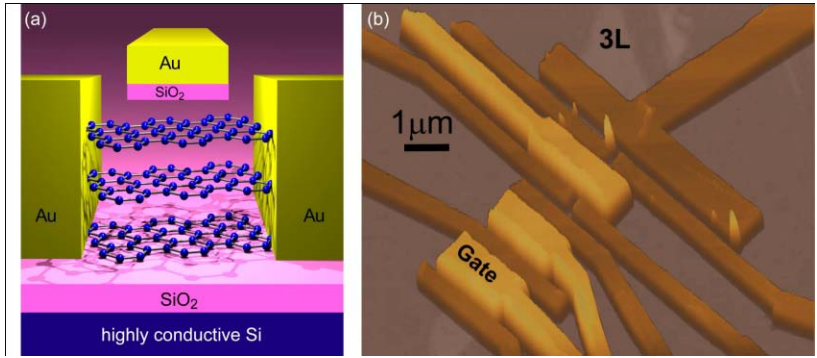

**Figure 4.** (a) Schematic structure of a double gated trilayer graphene device. The bottom gate is the Si substrate (dark blue) covered by $SiO_2$ (pink). The top gate is formed by $SiO_2$ (pink) and Au (yellow). The trilayer graphene is contacted by Au electrodes (yellow). (b) AFM image of graphene with top gates (light brown) and electrodes (dark brown).

A simple way to realize top gated graphene structures is to spin-coat a thin layer (~40nm) of polymethylmethacrylate (PMMA) to act as the dielectric material and then cover it with a metal electrode [22, 38, 78]. Exposure of PMMA to a high dose of electrons (21 mC/cm$^2$) cross-links the PMMA molecules (effectively forming a network of larger molecules), making it robust to solvents that would otherwise dissolve it. The top metal-gate electrode is then fabricated by standard electron-beam lithography and lift-off on top of the cross-linked PMMA layer. Top gates fabricated by this technique allowed a maximum charge carrier modulation in graphene of $5*10^{12}$cm$^{-2}$ [38] and they have been successfully employed to investigate electronic properties of *p-n* junctions in single layer graphene. However, a higher charge density modulation and a large perpendicular electric field applied onto the graphene-materials can only be achieved by means of dielectric materials with higher dielectric constant ($\varepsilon$) and a smaller thickness than the ~40 nm PMMA of Ref. [38].

The emerging technology of Atomic Layer Deposition (ALD) offers a valid alternative to the problem of growing thin dielectric materials with good insulating properties (e.g. low leakage current and high breakdown field). ALD allows the controlled deposition of one atomic layer at a time of almost any type of oxide dielectrics, including oxides with higher $\varepsilon$ than PMMA and $SiO_2$, see Table 1. This technique requires the deposition of a catalyst layer onto graphene suitable for the formation of the desired oxide –typically $Al_2O_3$ [30, 37, 80, 81] and $HfO_2$ [23, 34, 35, 83-87]. Although graphene is not chemically doped by this functionalization layer, in some cases the molecular structures attached to the graphene surface introduce a disorder potential on which the electrons in graphene scatter. Therefore, the electronic transport in graphene top gated structures defined by ALD is usually diffusive on a 100nm length-scale. The observation of physical phenomena which require ballistic electron transport –e.g. Klein tunneling- is only possible in very narrow top gates (less than 20nm in width [23]) which are technologically challenging to fabricate. Recent experiments showed that the use of a spin-coated organic seed layer made from a derivative of polyhydroxystyrene [34, 84] acts as a catalyst for the subsequent ALD growth of oxides while preserving the mobility of the underneath graphene layers.

| Top Gate dielectric | Dielectric constant / capacitance | Maximum charge density modulation achieved |
|---|---|---|
| Cross linked PMMA from **[38]** | $\varepsilon_{PMMA}$~4.5 | $n_{PMMA} \leq 5 \times 10^{12} cm^{-2}$ |
| ALD-deposited $Al_2O_3$ from **[37]** | $\varepsilon_{Al2O3}$~6 | $n_{Al2O3} \leq 2 \times 10^{12} cm^{-2}$ |
| hydrogen silsesquioxane (HSQ) + ALD-deposited $HfO_2$ from **[23]** | $\varepsilon_{HfO2}$~12 | $n_{HSQ+HfO2} \leq 4 \times 10^{12} cm^{-2}$ |
| Organic seed layer + ALD-deposited $HfO_2$, from **[34;86]** | $\varepsilon_{HfO2 + seed\ layer}$~10 | — |
| Electron-beam evaporated $SiO_2$ from **[33; 36]** | $\varepsilon_{SiO2}$~ 4 | $n_{SiO2}$ ~$1.1 \times 10^{13} cm^{-2}$ |
| Air-bridge **[21;40-42]** | $\varepsilon_{bridge}$~1 | $n_{bridge}$~$4 \times 10^{12} cm^{-2}$ |
| Polymer electrolyte from **[31]** (polyethylene oxide+LiClO$_4$) | $C_{polyethylene\ oxide+LiClO4}$ =1$\mu F/cm^2$ | $n_{polyethylene\ oxide+LiClO4}$~$5 \times 10^{13}\ cm^{-2}$ |
| Polymer electrolyte from **[92]** (polyethylene oxide+KClO$_4$) | $C_{polyethylene\ oxide+KClO4}$=7.4$\mu F/cm^2$ at 0.01Hz | $n_{polyethylene\ oxide+KClO4}$~$4 \times 10^{13}\ cm^{-2}$ |
| Ionic liquid from **[92]** | $C_{ionic\ liquid}$=120 $\mu F/cm^2$ at 0.01Hz | $n_{ionic\ liquid}$~$8 \times 10^{14} cm^{-2}$ |

**Table 1.** Dielectric constant, capacitance and maximum charge density modulation in top gate dielectric materials fabricated by various techniques.

A valid alternative fabrication method to ALD relies on the direct electron beam deposition of thin (15nm) SiO$_2$ on graphene and *in-situ* metallization of the gate [33]. In particular, it was experimentally demonstrated that the quality of electron-beam evaporated SiO$_2$ dielectric –e.g. breakdown field and leakage current- depends critically on the details of the metallization of the top gate electrode. For a high quality SiO$_2$ dielectric properties it is crucial that the oxide deposition is directly followed by an *in-situ* metallization of the top electrodes without exposure to air of the SiO$_2$ layer (Fig. 5a). This is evident when comparing the breakdown field and leakage current for top gates fabricated by *in-situ* deposition of SiO$_2$/metal interface -type A devices in Fig. 5- and by exposing the SiO$_2$/metal interface to air –type B devices in Fig. 5a and b. Respectively, type A devices show higher and reproducible values of breakdown fields compared to type B, and they are comparable to that of thermally grown SiO$_2$ (Fig. 5c). A clear advantage offered by electron beam deposited SiO$_2$ over the previously reviewed gate fabrication techniques is the possibility to achieve much higher values of external perpendicular electric fields applied onto the graphene-materials. This is evident when considering that the typical SiO$_2$ thickness is 10-15nm against the 20-30nm of the ALD bilayer dielectric and the 40nm of cross-linked PMMA. Therefore, electron beam deposited SiO$_2$ gate dielectrics are preferred to investigate the gate tunability of FLG such as a band-gap opening in bilayer graphene.

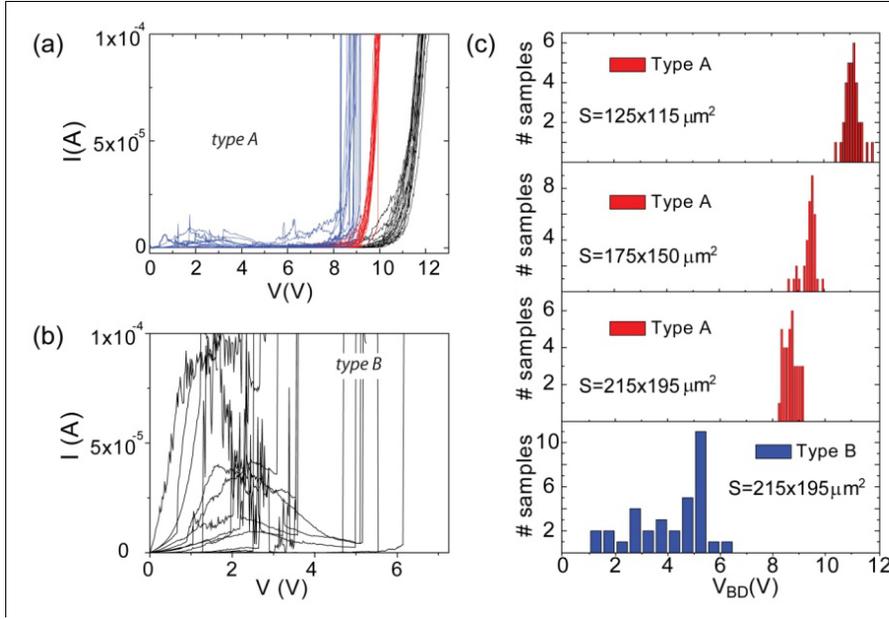

**Figure 5.** (a) Leakage current plotted as a function of gate voltage for *type A* top gates – fabricated by *in-situ* deposition of $SiO_2$/metal interface– with different areas: $S = 215\times195\mu m^2$ (blue curves), $S=175\times150\mu m^2$ (red curves) and $S = 125\times115\mu m^2$ (black curves). (b) *Type B* top gates –fabricated by exposing the SiO2 metal interface to air– with $S = 215\times195\mu m^2$ show higher leakage currents than *type A* top gates. The fact that *in-situ* deposition gives the best quality top gated structures is also demonstrated by the histograms in (c) of the breakdown voltage $V_{BD}$ for *type A* devices (with different areas) and for *type B* devices. For a fixed surface area (215×195μm2), the breakdown voltage $V_{BD}$ for *type A* is typically in the range $8\ V < V_{BD} < 9\ V$ whereas *type B* devices break down anywhere in the range $0\ V < V_{BD} < 6\ V$. The increase of $V_{BD}$ with decreasing the area for *type A* devices possibly indicates that the properties of $SiO_2$ close to breakdown are determined by small defects present in the film. Reprinted with permission from [33]. Copyright 2009 IOP Publishing Ltd.

The fabrication methods described above rely on embedding graphene in a sandwich structure between oxide dielectrics. These geometries do not allow post-fabrication annealing commonly used to remove resist residuals and adsorbants from the graphene surface. "Air-bridge"-styled top gates overcome this problem by suspending a metallic bridge across the graphene flake, with air or vacuum acting as the dielectric (see Table 1) [21, 40-42]. Fig. 6a illustrates the procedure for the fabrication of suspended top gates. Two layers of PMMA with different molecular weights are spun on the flake: a soft resist (495 K) on top of a hard resist (950 K). Patterning is accomplished via low-energy (10 kV) e-beam lithography and different exposure doses for the bridge span and pillars. The dose in the span is just enough to expose the soft resist but too small to affect the underlying hard layer. Both layers are exposed at a larger dose in the areas of the pillars (and contacts). The structures are then developed and covered with 5/250 nm of Cr/Au. A standard lift-off procedure removes PMMA leaving the bridge with a span up to 2 μm supported by 2 pillars (see Fig. 6b).

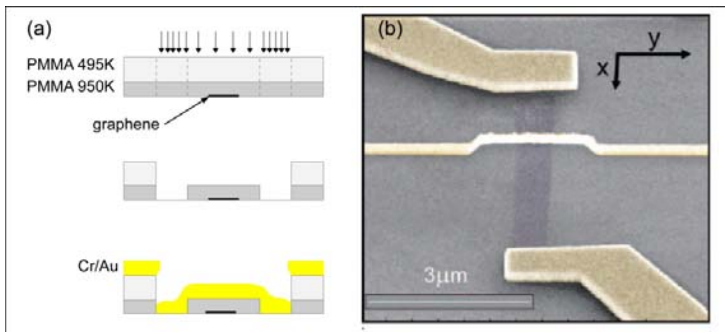

**Figure 6.** (a) Three stages of the air-bridge fabrication. (b) A false-colour SEM image of a graphene flake with a metal air-bridge gate. Reprinted with permission from Ref. [21]. Copyright 2008 American Chemical Society.

The need to combine electron transport with optical studies [30, 31] requires optical access to the graphene material underneath the gate electrode. Transparent top gates have been fabricated by deposition of $Al_2O_3$ dielectric and a very thin Pt film (20nm) that is electrically conductive but optically transparent [30]. Another method for the fabrication of top-gate structures compatible with optical access to the graphene is based on transparent solid polymer electrolyte (where inorganic ions are solvated into an organic polymer) [31]. In this case the solid polymer electrolyte covers the graphene flake whereas the metal gate electrode is placed next to the flake, rending the top gate fully transparent. A voltage applied to the metal gate electrode creates mobile ions in the electrolyte. These ions are driven to the graphene/electrolyte interface where they form an electronic double layer (EDL) acting as a capacitor.

The ionic gating of graphene by means of solid polymers not only allows for transparent gates, but the capacitance of such a gate (~1 µF/ $cm^2$) is two orders of magnitude larger than that of gates based on oxide dielectrics (~12nF/$cm^2$ for 300nm $SiO_2$) because the EDL is only a few nm thick. Accordingly, an ionic gate can accumulate more carriers and give access to much greater electric field strengths than conventional, oxide-based gates (see Table 1). Such a high carrier density is valuable for technological applications (e.g. low voltage graphene transistor operation [88] and high-sensitivity graphene-based chemical and biological sensors [89]), as well as for observing new physical phenomena predicted to occur in the high carrier density regime (e.g. superconductivity [90, 91]). In this respect, ionic liquids (composed only of organic ions) enable even higher ion concentration than polymer electrolytes, resulting in a further reduction of the EDL thickness (~ 1nm) and higher charge density (approaching $n_{ionic\ liquid} \sim 10^{15} cm^{-2}$ [92]).

## *p-n* junctions in single layer graphene

The ability to control locally the charge type and density in top gated graphene devices enabled a conceptually novel way to fabricate *p-n* junctions. Unlike in standard semiconductors where the carrier type is fixed by *chemical doping* during the growth process, in graphene the Fermi level can be continuously driven between the valence and conduction bands simply by applying a gate voltage, i.e. *electrostatic doping* which leads to ambipolar transistors. Therefore, local top gates fabricated on graphene allow the independent control of the electric field polarity applied to adjacent graphene regions –e.g. one covered by a top gate and the other not, originating an in-plane gate tunable *p-n* junction [21-23, 37-42]. Although graphene is certainly not the only material to exhibit this ambipolar transistor behaviour (e.g. organic single crystals [93, 94]), the high charge carriers mobility (higher than in silicon) makes graphene the ideal platform for investigating experimentally a large variety of physical phenomena both in the most common diffusive transport regime as well as in the ballistic regime.

In what follows we will review the charge transport through in-plane gate tunable *p-n* junctions in single-layer graphene. At first we will summarize the electrical characteristics of diffusive junctions, and then we will discuss the feature of novel physical phenomena, such as Klein tunneling, characteristic of ballistic Dirac particles impinging on a tunnel barrier.

### *Diffusive junctions*

Fig. 7a shows the simplest form of a single graphene *p-n* junction [37]: the back gate is used to control the charge density over the whole flake (region1 and region2), whereas the top gate controls the charges in the region below it (region2). The independent control of carrier type and density in the two regions (i.e. Fermi level) is demonstrated by measuring the resistance across two contacts, one outside the top gate region and one underneath the top gate, as a function of back gate voltage $V_{BG}$ and top gate voltage $V_{TG}$ (see Fig. 7b). A color scale plot of the sheet resistance reveals a skewed ridge corresponding to the maximum value of resistance (charge–neutrality point). This divides the ($V_{TG}$,$V_{BG}$) space into four quadrants with distinct carrier type (holes or electrons) for each of the two regions forming the sample. The charge neutrality point in region 1 does not depend on $V_{TG}$, and therefore appears as a horizontal ridge. On the other hand, the charge neutrality in region 2 depends on both top and back gate forming a diagonal ridge in the plot of Fig. 7b. At ($V_{TG}$,$V_{BG}$) ≈ (−0.2 V, −2.5 V) charge neutrality is attained throughout the entire graphene sample.

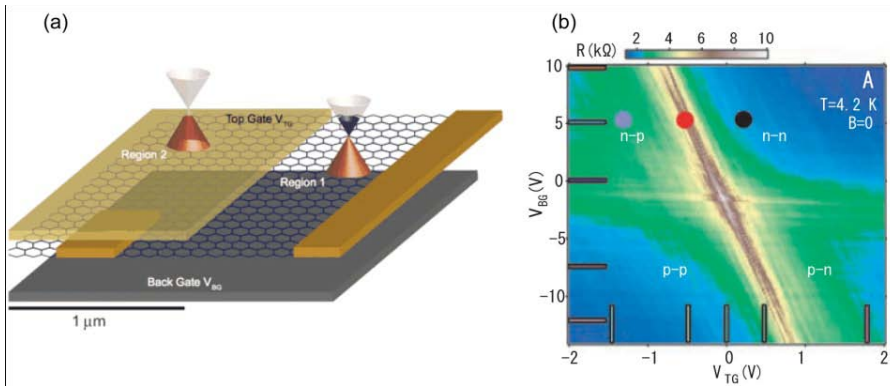

**Figure 7**. Panel (a) shows a schematic design of *p-n* junction device realized on a single layer graphene. The color coded plot in panel (b) shows the two-terminal resistance as a function of $V_{TG}$ and $V_{BG}$ measured between the contacts in Region1 and Region2. These measurements demonstrate the independent control of carrier type and density in both region 1 and 2. (From ref. [37]. Reprinted with permission from AAAS. Copyright 2007 by the American Association for the Advancement of Science.)

Charge transport measurements of a *p-n* junction system in the quantum Hall regime reveal a rich scenario of quantization values of the conductance (Fig. 8). For a single layer graphene flake with uniform carrier density and type, the quantum Hall effect manifests in a unique series of conductance plateaus at half integer multiples of $4e^2/h$ [3,4], as introduced in the section *electronic properties of graphene materials*. This integer QHE in graphene is responsible for the plateaus at $6e^2/h$ and $2e^2/h$ observed in the unipolar regime by Williams *et al.* [37].

However, the measurements in Fig. 8 clearly demonstrate the emergence of new plateaus at $e^2/h$ and $(3/2)e^2/h$. The origin of these plateaus is traced back to the equilibration of edge states at the *p-n* interface, and are characteristic of this in-plane heterointerface as theoretically calculated by D.A. Abanin *et al.* [95]. When the filling factors $\nu_1$ and $\nu_2$ of region 1 and 2 have the same sign (n-n or p-p case), the edge states common to both regions propagate from source to drain (see Fig. 8d), whereas the remaining $|\nu_1 - \nu_2|$ edge states in the region of highest absolute filling factor circulate internally within that region without contributing to the conductance. This gives rise to conductance plateaus at g=min($|\nu_1|$,$|\nu_2|$)x$2e^2/h$. On the other hand, if the filling factors $\nu_1$ and $\nu_2$ have opposite sign (*n-p* or *p-n*), the counter-circulating edge states in region 1 and 2 travel in the same direction along the *p-n* interface (see Fig. 8e). This facilitates mode mixing between parallel travelling edge states. Whenever the current entering the junction region is uniformly distributed among the parallel travelling modes, complete mode mixing occurs and quantized plateaus are expected at g=$|\nu_1||\nu_2|/(|\nu_1|+|\nu_2|)2e^2/h$.

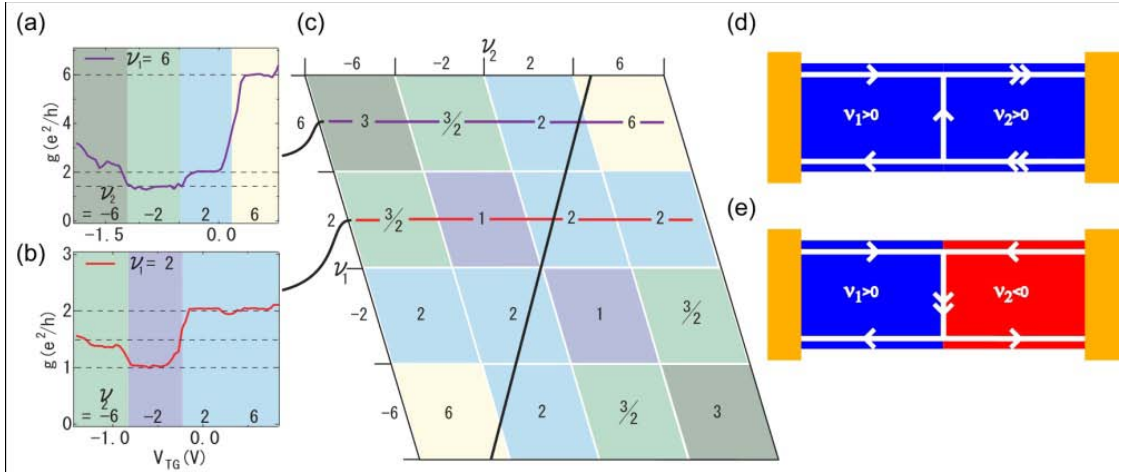

**Figure 8.** Panel (a) and (b) show conductance measurements respectively at $\nu_1 = 6$ and $\nu_1 = 2$. The conductance displays plateaus at 6, 2, and $(3/2)$ $e^2/h$ in (a) and plateaus at 2, 1, and $(3/2)$ $e^2/h$ in (b). The table in (c) summarizes the expected conductance plateau values as a function of filling factors. The purple and red lines correspond to slices in (a) and (b). (From ref. [37]. Reprinted with permission from AAAS. Copyright 2007 by the American Association for the Advancement of Science). (d) Schematic of edge states propagating from source to drain in regions with filling factors of the same sign. (e) Schematic of counter-circulating edge states in regions with filling factors of opposite sign.

The single *p-n* graphene junction represents only the first step towards more complex in-plane heterostructures with novel electronic functionalities as compared to standard semiconducting devices. For instance, two back to back *p-n* junction devices form an in-plane bipolar junction device with gate tunable potential barriers [38-42]. Graphene *p-n-p* (*n-p-n*) structures were first experimentally realized by Huard *et al.* [38], see Fig. 9. The combination of a back and top gates identifies 3 distinct graphene regions whose charge densities $n_2$ (underneath the top gate), $n_1=n_3$ (everywhere else) and charge carrier type are controlled by the gate voltages. Clear evidence for the presence of these 3 distinct graphene regions is found in two probe resistance ($R_{2p}$) measurements of the device as a function of $V_{BG}$ and $V_{TG}$ (see Fig. 9b). $R_{2p}$ displays maximum values for specific combinations of gate voltages applied to the top and back gate -i.e. white ridges in the colour code plot of Fig. 9b. When the Fermi level lays on the charge neutrality point in region 1 and 3, $R_{2p}$ exhibits a top gate independent ridge; *vice versa* when the Fermi level lays on the charge neutrality point in region 2, the ridge of $R_{2p}$ is skewed in diagonal because it is affected by a combination of $V_{BG}$ and $V_{TG}$.

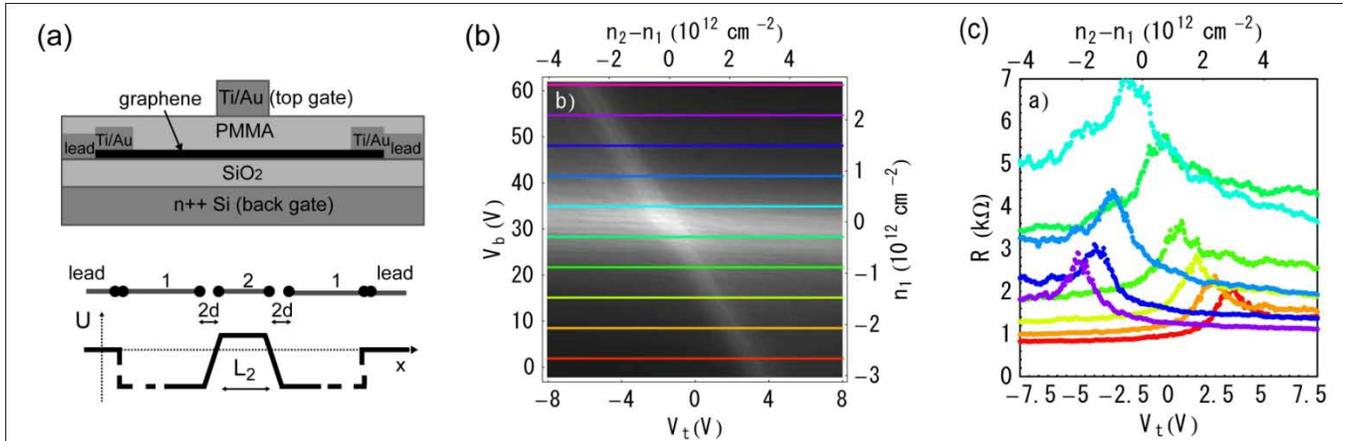

**Figure 9**. (a) Cross-section view of a double gated bipolar junction with cross-linked PMMA as a top-gate dielectric. Bottom panel shows a simplified model for the electrochemical potential U of electrons in graphene along the cross section of the device shown on top. The potential is shifted in region 2 by the top gate voltage and shifted in both regions 1 and 2 by the back gate voltage. (b) Two-dimensional gray scale plot of the resistance across the graphene sample at T=4K as a function of both gate voltages. Here $V_b$ is the back gate voltage (reffered to in the main text as $V_{BG}$) and $V_t$ is the top gate voltage ($V_{TG}$). The resistance values are indicated in the curves of panel (c) which are cuts along the correspondingly colored lines of (b). Reprinted figure with permission from [38]. Copyright 2007 by the American Physical Society.

Transport measurements in these bipolar *p-n-p/n-p-n* junctions in the quantum Hall regime [39, 41, 42] reveal a highly complex scenario of new quantization values for the conductance, which differ significantly from those found for single *p-n* junctions [37]. Fig. 10 shows the conductance plateaus occurring at values close to fractional values of $e^2/h$, including *2/3, 6/7,* and *10/9*. These values of conductance quantization are characteristic of bipolar junctions and go clearly beyond the half integer quantum hall plateaus of $4e^2/h$ that are the hallmark of a homogeneous single layer graphene device. The understanding of the QHE in bipolar graphene junctions requires appropriate consideration of edge state formation and propagation together with the details of the charge density landscape throughout the device.

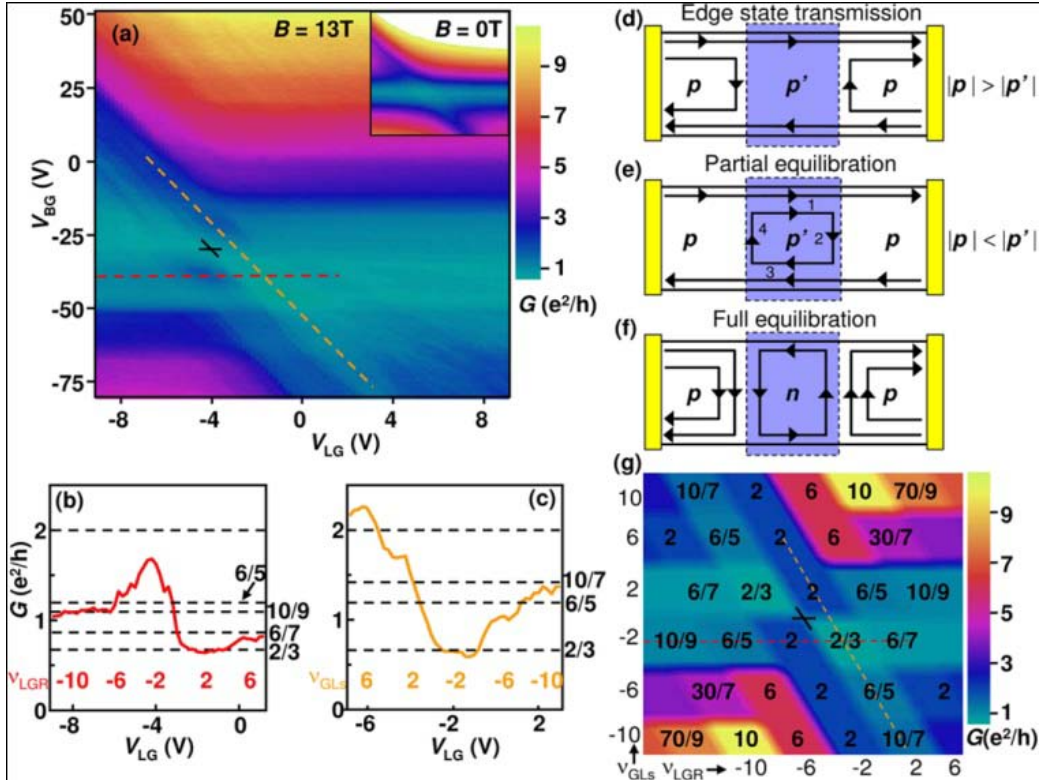

**Figure 10.** Panel (a) shows a colour code plot of conductance versus local top gate voltage ($V_{LG}$) and back gate ($V_{BG}$) at magnetic field B=13 T, and T=4.2 K. Here $V_{LG}$ represents the top gate voltage, reffered to in the main text as $V_{TG}$. The black cross indicates the location of filling factor zero. Inset: Conductance at zero B in the same range of gates and the same colour scale as main. Graphs (b,c) show the projection of the conductance traces extracted from (a) respectively along the red and orange dashed line, showing fractional values of the conductance. Numbers on the right indicate expected fractions for the various filling factors -numbers below the trace indicate the filling factor; see also the simulated colour map of the theoretical conductance plateaus in (g). (d), (e), and (f) show different edge state diagrams representing possible equilibration processes taking place at different charge densities. Reprinted figure with permission from [39]. Copyright 2007 by the American Physical Society.

We start from considering a larger number of edge-states outside the top gate than under it (Fig. 10d) –that is the same carrier type is accumulated throughout the device with lower charge density under the top gate than outside (e.g. $n$-$n_0$-$n$ or $p$-$p_0$-$p$). In this case, the edge-states existing only outside the top gate are fully reflected at the *p-n* interface while those present in both regions are fully transmitted. Naturally only the transmitted edge-states contribute to the net conduction throughout the device. A more interesting situation –unique to bipolar junctions- occurs when the density under the gate is higher than the density outside the gate, with the same carrier type over the whole device (Fig. 10e). In this case the large number of edge-states under the local gate can partially equilibrate the different edge-states outside the local gate, producing the quantization sequence $G(e^2/h)$= 6/5, 10/9, 30/7... [39, 41, 42]. Conversely, when the regions outside and under the local gate have opposite carrier type accumulation, the edge-states counter-circulate in the *p* and *n* areas (see Fig. 10f), running parallel to each other along the *p-n* interface. Such propagation leads to mixing and full equilibration among edge-states at the *p-n* interfaces, with conductance plateaus at $G(e^2/h)= 2/3, 6/5, 6/7...$ [39, 41, 42].

*Ballistic junctions*

The opposite transport regime to the diffusive is the ballistic regime, that is charge carriers can travel throughout a *p-n-p* (*n-p-n*) bipolar junction device without experiencing scattering events [21-23]. This regime hosts a variety of novel physical phenomena which highlight the massless relativistic character of charge carriers in graphene. Klein tunneling [19, 20], particle collimation [23], and Veselago lensing [24] are physical phenomena intimately related to ballistic Dirac fermions. In this section we will review the main experimental evidence for ballistic transport in bipolar graphene junctions and its implications.

    The Klein paradox [96] predicts that an incident relativistic electron on an infinite potential barrier transforms into an anti-particle (positron) inside the barrier and moves freely in this otherwise forbidden region, i.e. high transmission probability through an infinite barrier [19, 20]. Thanks to the relativistic nature of charge carriers in graphene, it is now possible to observe the Klein paradox on a chip. We consider a normal incident charge carrier to a graphene *n-p* interface with a smooth potential which does not introduces inter-valley scattering [46, 47]. Due to the chiral nature of Dirac fermions in graphene, for an incoming electron in the *K* point the pseudospin is parallel to the momentum. Since the *n-p* interface potential is smooth on the atomic scale, the pseudospin is conserved in the tunneling process [46, 47]. Therefore, the electron impinging perpendicular onto the *n-p* junction cannot backscatter in a counter propagating electron state which would be in the *K'* point, but it scatters in a counter propagating hole state which has opposite momentum direction but with the same pseudospin direction as the impinging electron. This leads to a transmission probability 1 through the *p-n* junction.

    More in general, due to Klein tunneling normally incident charge carriers are perfectly transmitted through a ballistic *p-n* junction whereas obliquely incident charge carriers have a transmission probability which depends on the angle of incidence. This transmission of carriers through graphene *p-n* junctions resembles optical refraction at the surface of metamaterials with negative refractive index: i.e. the electric current is focussed by a graphene *p-n* junction. This effect can be extended to realize electric current lenses in *n-p-n* structures. Furthermore, if we consider particles incident onto such bipolar *n-p-n* (*p-n-p*) graphene junctions under angles where neither the transmission probability nor the reflection probability are too large, quantum interference phenomena have to be expected since a resonant cavity is formed between the two p-n interfaces.

    The first experimental evidence of Klein tunneling in graphene-based ballistic bipolar junctions was reported in electric transport measurements in double-gated graphene devices with a bridge gate electrode [21]. In comparison to the diffusive transport regime, the maximum resistance measured in the ballistic regime through bipolar junctions is systematically higher (see Fig. 11). This can be understood when considering that due to the Klein tunneling process, the tunneling probability for obliquely impinging electrons to each *p-n* junction is lower in the ballistic transport regime than in the diffusive case. Therefore a higher interface resistance has to be expected for the ballistic regime compared to the diffusive regime. In addition, experiments revealed reproducible oscillations of the resistance maximum as a function of the gate applied to the bridge electrode (Fig. 11b, c and d). These oscillations are intimately related to oscillations of the transmission coefficient caused by quantum interference of chiral carriers within a ballistic bipolar junction [21]. Similar transport experiments in double gated structures provided evidence of Klein tunneling in the quasi-ballistic regime [22] achieved at high charge density induced by the top gate ($n_{TG}>3\times10^{12}cm^{-2}$). However, the direct observation of quantum interference effects in these devices requires fully ballistic transport throughout the bipolar junctions.

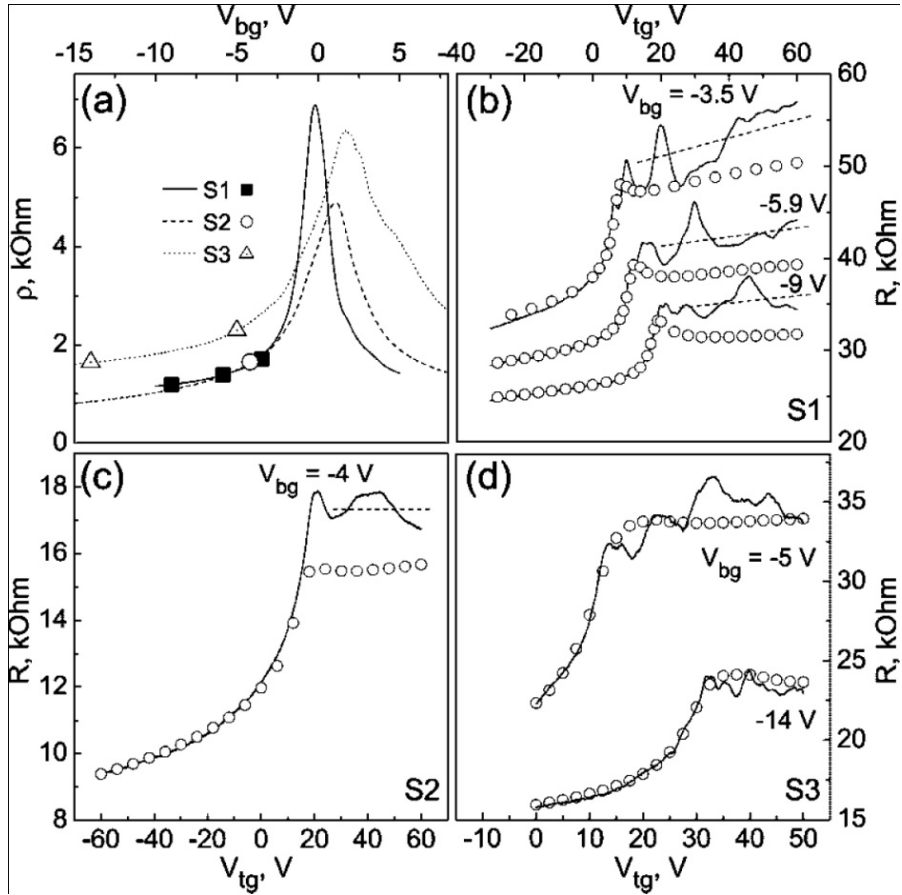

**Figure 11.** (a) Resistivity of samples with different mobility as a function of the back-gate voltage, at Vtg= 0. Here $V_{bg}$ is the back gate voltage (reffered to in the main text as $V_{BG}$) and $V_{tg}$ is the top gate voltage ($V_{TG}$). Points indicate the fixed values of Vbg where the top-gate voltage was swept to produce *p-n-p* junctions. (b) The resistance of sample S1 as a function of top gate voltage at different Vbg. (c,d) The resistance as a function of top-gate voltage at different Vbg of samples S2 and S3, respectively. Points show the results of the calculations of the expected resistance assuming diffusive transport of carriers. Reprinted with permission from [21]. Copyright 2008 American Chemical Society

Magneto-transport experiments focussing on this oscillatory part of the resistance were able to estimate the magnitude and phase of the transmission and reflection coefficients through a ballistic heterojunction [23]. A perpendicular magnetic field (*B*) applied to the graphene bends the trajectories of the carriers, and therefore it modifies the charge carriers' angle of incidence at each *p-n* junction. Since the transmission (and reflection) coefficient is a function of the angle of incidence at the *p-n* junction (Klein tunneling), the external magnetic field allows the control of transmission and reflection coefficients in graphene *p-n* junctions. As *B* increases, the cyclotron bending favours the transmission of modes incident on the junctions at angles with the same magnitude but opposite sign. For perfect transmission, at normal incidence, the reflection amplitude changes sign as the sign of the incidence angle changes, causing a π shift in the phase of the reflection amplitudes [97]. In the devices investigated by Young *et al.* [23] the field at which this phase shift occurs is in the range of 250-500mT (see Fig. 12).

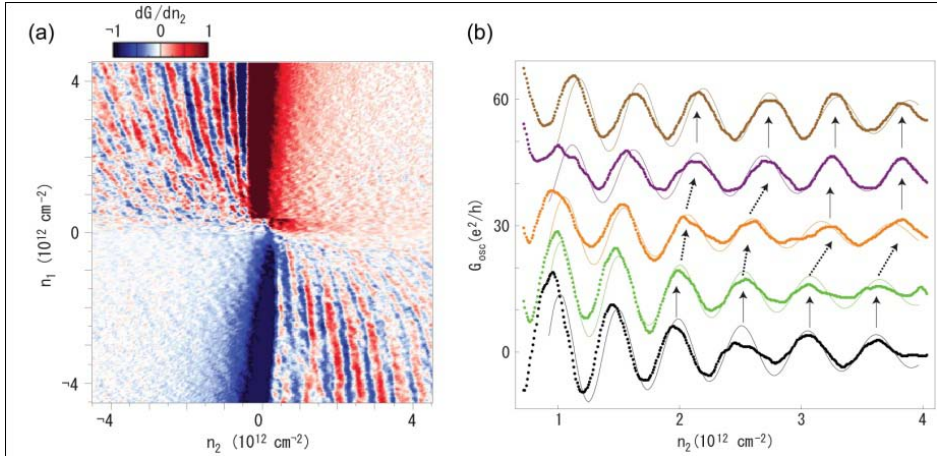

**Figure 12**. Panel (a) shows a color coded plot of the derivative of the conductance with respect to the charge density induced by the top gate ($n_2$) as a function of top ($n_2$) and back ($n_1$) gate charge densities forming bipolar p-n junctions. The oscilations in conductance are aparent when the densities $n_1$ and $n_2$ have opposite signes. (b) Conductance measurements *vs.* $n_2$ for fixed $n_1$ at different external magnetic field values applied perpendicular to the sample (from the bottom to the top conductance curve B=0, 200, 400, 600 and 800mT); the dots represent data, the smooth lines are the result of the simulations. The sudden phase shift that signals the presence of perfect transmission is indicated by dotted arrows. Curves are offset for clarity. (From ref. [23]. Copyright 2009 Macmillan Publishers Limited.)

**Electric field tayloring of few layer graphene's band structure**

Although the ability to control locally the charge type and density in single-layer graphene has allowed the investigation of unprecedented physical phenomena, graphene-transistors lack of a necessary requirement for nowadays applications. The gapless low energy band dispersion of graphene – governed by the equivalence between *A* and *B* lattice sites, see Fig. 1- makes it difficult to envisage transistor devices in which a gate voltage can switch on and off the flow of charge carriers. So far, various approaches have been proposed or implemented to break the symmetry between the *A* and *B* sites leading to the opening of an energy band gap in graphene. Examples are offered by strain engineering [98, 99], graphene-substrate interaction [100, 101], confinement [102] and chemical modifications of graphene [103].

An alternative solution to the problem of creating a band gap in graphene is offered by bilayers [25-35, 48, 104]. In this case, the equivalence between the $A_1$ and $B_2$ sites (see Fig. 1) makes bilayers zero gap semiconductors with parabolic energy bands touching at *E=0*. Breaking the symmetry between the two atomic layers –i.e. rendering the $A_1$ and $B_2$ sites energetically inequivalent- shifts the low energy parabolic bands to higher energies and opens a band-gap (*Δ*). Since $A_1$ and $B_2$ belong to different layers, we can now use the dual valency of double-gated devices to control the Fermi level in the flakes and the induced electrostatic potential energy asymmetry between the two layers. This electric field induced asymmetry leads to an electric field tunable band gap [48, 104]. Once the energetic equivalence between the $A_1$ and $B_2$ sites is restored, the energy gap reduces to zero and the semimetallic character of bilayers is recovered. Although the breaking of the interlayer symmetry was experimentally realized in various ways (e.g. charge transfer from the substrate to epitaxial graphene [25], chemical doping of mechanically exfoliated graphene on $SiO_2$ [26]), the double gated device architecture remains most attractive for real life electronic applications since it offers a straightforward way to independently and continuously control *in-situ* both the band gap and the Fermi level by means of gate voltages [27, 30, 31, 33-35].

The opening of an energy band gap in bilayer graphene has a characteristic fingerprint in the magnetotransport measurements in Hall bar devices (Fig. 13) [26]. The quantum Hall sequence of the conductivity characteristic of a zero-gap bilayer graphene follows the integer sequence $\sigma_{xy}=\pm(4e^2/h)N$ for $N \geq 1$ with no plateau at $N=0$, consistent with a metallic state at the neutrality point (see Fig. 13c [26]). When a gap opens up in the energy band of bilayers a plateau appears at N=0 in the Hall conductivity (see Fig. 13c). The equidistant plateaus of $\sigma_{xy}$ are the hallmark of the quantum Hall effect for an ambipolar semiconductor with an energy gap exceeding the cyclotron energy. The actual value of the gap can be extracted from a tight binding model, where the relevant hopping parameters are estimated from a fit to the experimental magnetotransport data (Shubnikov-de Haas oscillations). This analysis shows that a mid-infrared energy gap opens up in the band structure of bilayers by using perpendicular electric fields smaller than 1 V/nm.

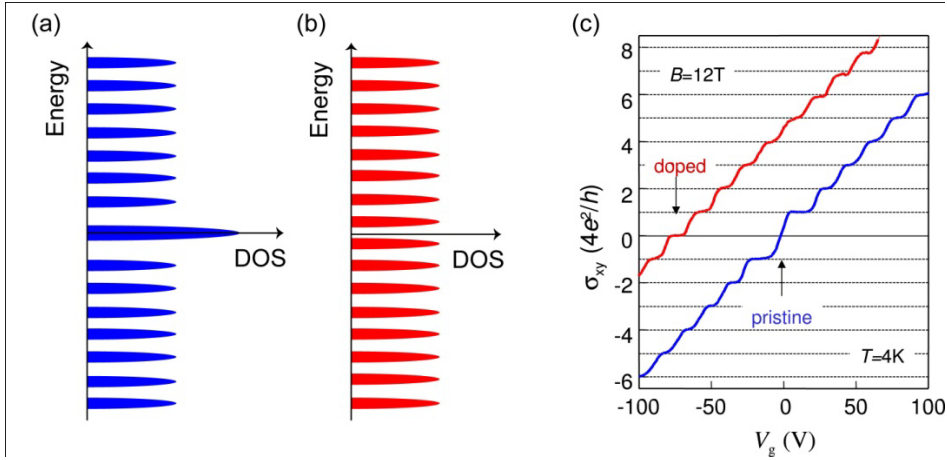

**Figure 13.** The sequence of Landau levels in the density of states (DOS) of bilayer graphene with two degenerate levels $N = 0$ and $N = 1$ at zero energy (a) and of bilayer graphene with the gap opened, where the degeneracy of the zero-energy Landau level is lifted (b). (c) Quantum Hall Effect in doped and pristine bilayer graphene. In pristine bilayer, $\sigma_{xy}$ exhibits plateaus at all integer $N$ of $4e2/h$ except for $N = 0$. The $N=0$ plateau is recovered in chemically doped bilayer when the gap is opened. The doping also shifts the neutrality point to high $V_g$. Reprinted with permission from [26]. Copyright 2007 by the American Physical Society

We now turn to discuss experiments on double gated bilayer graphene structures [27, 33]. A finite voltage applied to either of the gates (back or top gate) changes the position of the Fermi level in the gated region of the graphene by an amount corresponding to the induced charge density. When the two gates are biased with opposite polarity, a large external electric field ($E_{ex}$) applied perpendicular to the layer is generated: $E_{ex}=(V_{BG}-V_{TG})/(d_{BG}+d_{TG})$ with $d_{BG}$ and $d_{TG}$ the thicknesses of, respectively, the back and top dielectric. In this device configuration, the evolution of the in-plane transport properties shows evidence for the opening of a band gap for $E_{ex}\neq 0$, as we will discuss in the following.

Fig. 14a shows typical electric transport measurements of the in-plane bilayer graphene square resistance ($R_{sq}$) as a function of the voltage applied to the back gate, with the top gate at a constant potential. The resistance clearly exhibits a maximum ($R_{sq}^{max}$) whose value and position in back gate voltage depends on the voltage applied to the top gate. In particular, $R_{sq}^{max}$ increases for progressively larger external electric fields and its value for $E_{ex} \neq 0\ Vm^{-1}$ exhibits a pronounced temperature dependence typical of an insulating state (see Fig 14b) – i.e. an electric field induced energy gap has opened. In the absence of disorder, when a gap opens in the band structure of bilayer graphene, the density of states in the gapped region is zero. In this ideal case, the value of $R_{sq}^{max}$ is defined by thermally activated charge carriers ($R_{sq}^{max} \propto e^{\Delta/2k_BT}$), and $\Delta$ can be accurately determined from the temperature dependence of $R_{sq}^{max}$. However, transport measurements in double gated bilayer devices consistently revealed that $R_{sq}^{max} \propto \exp(T_0/T)^{1/3}$ with $T_0$ a fitting parameter dependent on the disorder in the sample. This specific temperature dependence is characteristic of variable-range hopping for non-interacting two-dimensional charge carriers in an insulator, and with disorder induced sub-gap density of states assisting the transport. This variable-range hopping hinders a direct estimate of $\Delta$ in transport experiments.

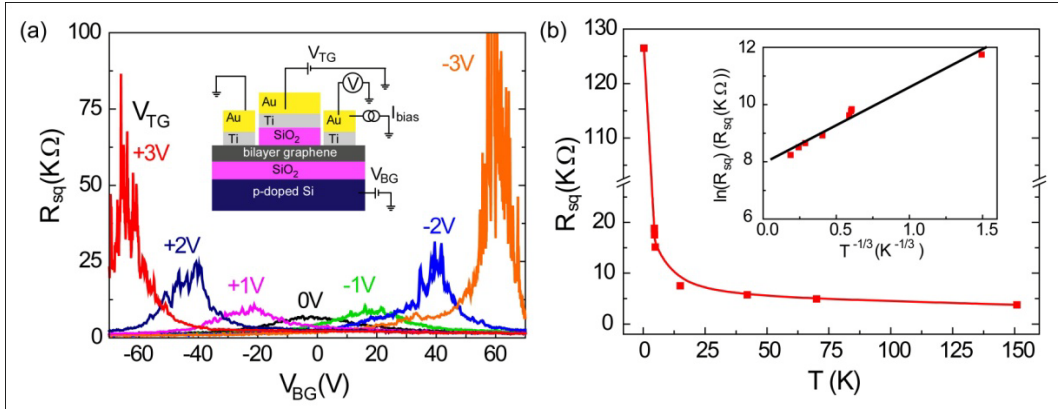

**Figure 14.** (a) Square resistance $R_{sq}$ of bilayer graphene as a function of back gate voltage ($V_{BG}$) measured for different fixed values of the top gate voltage ($V_{TG}$) at $T=300mK$. The inset shows the double gated bilayer graphene device configuration. The position of the Fermi level and the applied perpendicular electric field are controlled by $V_{BG}$ and $V_{TG}$. The resistance is measured by applying an ac-current bias $I_{bias}$ and measuring the resulting voltage V across the device. (b) Temperature dependence of $R_{sq}$ and $ln(R_{sq})$ (inset) for $V_{BG}= -50V$ and $V_{TG}=3V$. Reprinted with permission from [33]. Copyright 2009 IOP Publishing Ltd.

Possibly the most direct measurement of an electric field induced band gap in double gated bilayer graphene comes from infrared spectroscopy experiments [28-31]. Whenever a band-gap is open in bilayer graphene, the infrared absorption displays a highly intense peak in the absorption spectra corresponding to the transition of charge particles from the top of the valence band to the bottom of the conduction band, see Fig. 15b. In particular the infrared absorption spectra shows a peak below 300meV –corresponding to the gap between valence and conduction band- with pronounced gate tunability: it gets stronger and shifts to higher energy with increasing the electric field. In this way, a continuously tunable band-gap from zero up to 250 meV has been directly observed, whereas the corresponding small increase of the maximum resistance as a function of $E_{ex}$ confirms a large disorder induced sub-gap density of states (Fig. 15a).

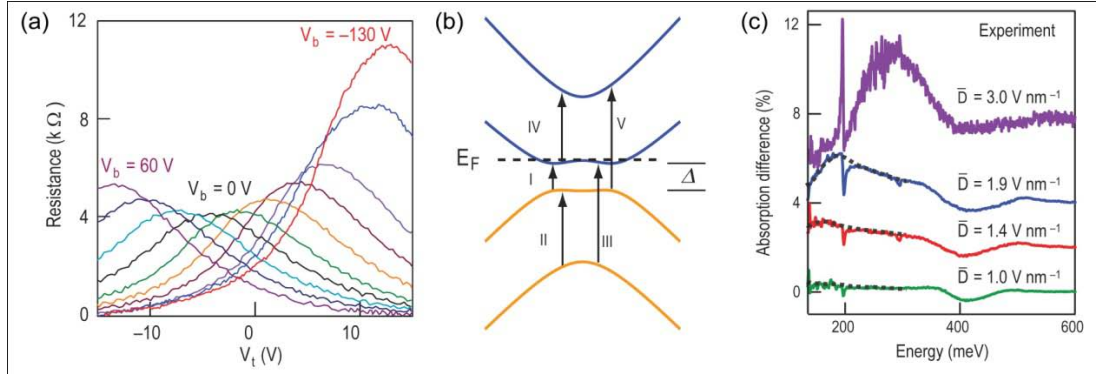

**Figure 15.** (a) Bilayer graphene electrical resistance as a function of top gate voltage ($V_t$) at different fixed back gate voltages ($V_b$). (b) Allowed optical transitions between different sub-bands of a graphene bilayer. (c) Gate-induced absorption spectra at the charge neutrality point for different applied perpendicular electric fields ($\bar{D}$). Absorption peaks due to transition I at gate-induced band gaps are apparent. The broad feature around 400 meV is due to electronic transitions II, III, IV and V. (From ref. [30]. Copyright 2009 Macmillan Publishers Limited.)

So far, transport experiments in double gated graphene bilayers revealed an on/off ratio of the current with values as high as 100 at room temperature for an average electric displacement of 2.2 V/nm [34], making bilayers an appealing material system for real life electronic applications. This gate tunable band-gap holds the promise for conceptually new devices such as, flexible and gate tunable light emitting devices and light detectors. However, the development of bilayer graphene applications is subdue to the improvement of the on/off ratio of the current in these structures. Disorder introduces states for sub-gap energies, therefore degrading the on/off current ratio [105]. Understanding and controlling this disorder is mandatory for the future development of bilayer-based electronics.

To date, single and double layer graphene are the most extensively studied materials belonging to the family of few layer graphene systems. Not much is known experimentally for thicker FLG and only recent experimental and theoretical attention has turned towards the properties of trilayer graphene. Fundamental questions, such as the evolution of the electronic properties from the one of mass-less Dirac electrons in a single layer to the massive particles of bulk graphite, can only be tackled by investigating the physics of few layer graphene systems which is so far largely unexplored. This knowledge gap prevents us from identifying the best suited thickness of FLG for a given application. For instance, the charge carriers mobility in bulk graphite is $10^6$ cm$^2$/Vs whereas typical mobilities measured in mechanically exfoliated single-layer graphene on SiO$_2$ are 20000 cm$^2$/Vs. Although naively one would expect a monotonous increase of mobility with increasing the number of layers, it was experimentally found that this is not the case –i.e. the charge carriers mobility decreases when increasing the number of FLG's layer up to 4 [36, 106].

Similarly, at the present it is not known experimentally the full extent of FLG' band structure gate-tunability. Bilayer graphene was found to be the only known material system with a gate-tunable band gap, but virtually nothing is known experimentally on thicker few layer graphene. Indeed, recent experiments by Craciun *et al.* [36] reported that, *ABA*-stacked trilayer graphene is the only known semimetal with an electric field tunable overlap between conduction and valence band. These experimental findings demonstrate that the rich physics of graphene materials extends well beyond the single and double layer to comprise the entire family of few layer graphene systems.

Magneto-transport experiments in double-gated FLG structures are a reliable and efficient tool to investigate the intrinsic band structure gate-tunability of these material systems. Fig. 16a shows a typical Hall coefficient ($R_H$) dependence on charge density –measured at fixed external perpendicular magnetic field- for a trilayer graphene. Noticeably, $R_H$ displays a characteristic sign reversal corresponding to the cross-over between different types of charge carriers in the conduction (electrons and holes). The back gate voltage range over which $R_H$ decreases and changes sign in trilayers is $\Delta V_{BG} \approx 8V$ and it is usually much larger than what is observed in single and double layer graphene devices. Such a large $\Delta V_{BG}$ suggests that trilayer graphene is a semimetal with a finite overlap between the conduction and valence bands ($\delta\varepsilon \sim 28$ meV). Note that the energy scale of the trilayer's overlap goes well beyond the few meV energy scale typical of disorder induced electron-hole puddles observed in single and double layers [107].

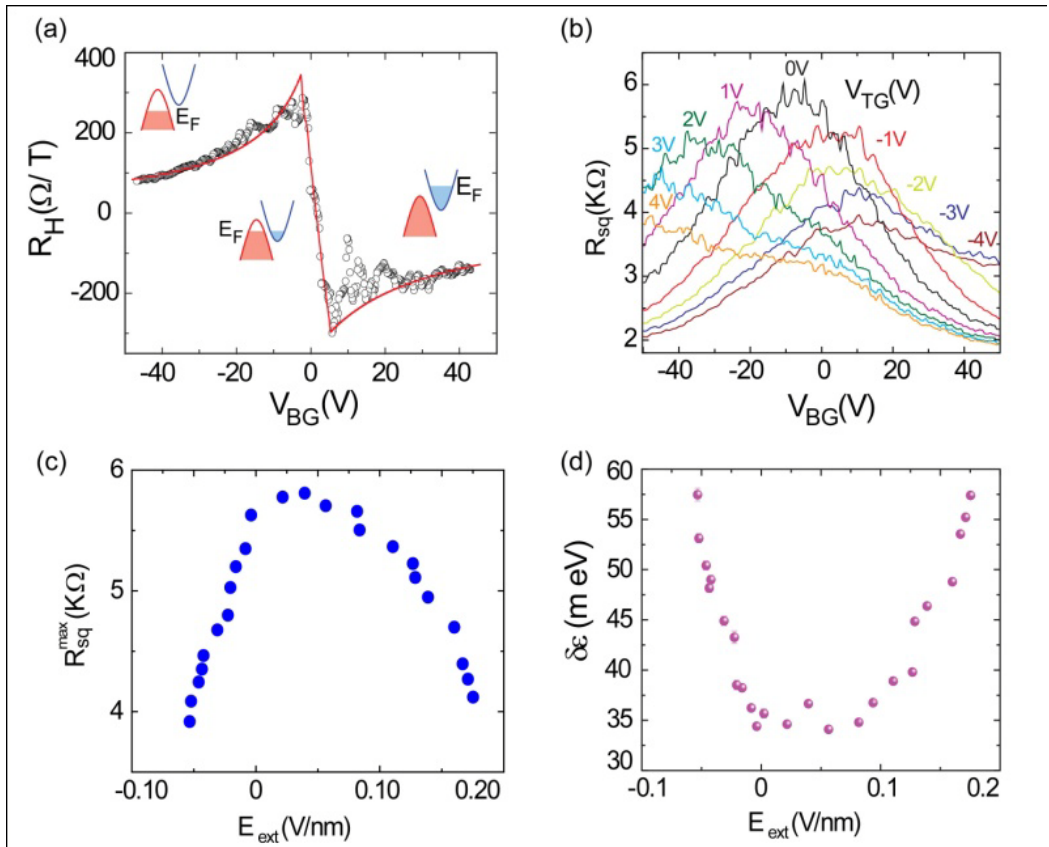

**Figure 16.** (a) Hall coefficient as a function of back gate voltage $V_{BG}$ (open circles) for a fixed perpendicular magnetic field of 9 T at 50 mK. The red curve is a fit. The three insets schematically depict the position of the Fermi level ($E_F$) at different values of $V_{BG}$. (b) Square resistance of a trilayer device versus $V_{BG}$ for different values of the top gate voltage $V_{TG}$ at 50 mK. (c) Maximum square resistance $R_{sq}^{max}$ versus external perpendicular electric field $E_{ex}$ at 50 mK. (d) Band overlap ($\delta\varepsilon$) dependence on $E_{ex}$. (From ref. [36]. Copyright 2009 Macmillan Publishers Limited.)

The unique electric field tunability of the trilayers band structure is revealed by transport experiments in double gated devices. Fig. 16b shows the resistance of trilayer graphene as a function of back gate voltage for different constant top gate voltages. It is apparent that the value of $R_{sq}^{max}$ decreases as the top and back gates are biased with voltages of opposite polarity and of increasing magnitude (i.e., $R_{sq}^{max}$ decreases when the applied perpendicular electric field increases, see Fig. 16c). The observed decrease of $R_{sq}^{max}$ as a function of external perpendicular electric field is opposed to the increase of $R_{sq}^{max}$ characteristic of bilayers (see Fig. 14). Systematic studies of $R_{sq}^{max}$ vs. $E_{ex}$ as a function of temperature reveal that the overall electric field dependence of the in-plane trilayer resistance is well explained within a two band model with an electric field tunable band overlap between conduction and valence band ($\delta\varepsilon$), see Fig. 16d. To this end trilayer is the only known semimetal with a gate tunable band overlap.

**Future directions**

The ability to nanofabricate single- and few-layer graphene double gated devices is enabling a paradigm shift in condensed matter, which benefits of the unique independent control of the Fermi level and the *in-situ* gate tunable band structure of these materials. In this review, we have summarized the gamut of novel physical phenomena and unprecedented material functionalities accessible in double gated graphene-based devices. Examples include the observation of relativistic tunneling of Dirac particles across graphene *p-n* heterointerfaces (i.e. Klein tunneling), the discoveries of a gate tunable energy band gap in bilayers' band structure and a gate tunable conduction-valence band overlap in trilayers' band structure.

Since the early stages of the development of double gated graphene devices, it was recognized that these structures hold the promise to bring solid state electronic devices a step closer to optics. In particular, the ability to control locally the charge density and type in single-layer graphene enabled the formation of in-plane gate tunable *p-n* junctions which are the right platform for investigating the tunneling of Dirac fermions. Theoretical proposals suggest that the angle dependent transmission (reflection) coefficient of Dirac fermions propagating throughout a graphene *p-n* interface, can be used to realize electron lenses on a chip. Due to Klein tunneling of relativistic electrons the electric current flowing in these devices can be focused simply by means of local gates which control the charge density and type, and without the need of an external magnetic field -so far the most common means used to bend the trajectories of charge carriers in a conductor [24]. This gate tunable electron lens is the fundamental device for more involved top gate shapes such as prism-shapes which can act as a focusing beam splitter for electrons. The ability to shape and eventually redirect by gate-refocusing a beam of Dirac particles in a graphene nano-circuit holds the promise for unprecedented device functionalities, and it is currently an open research field which is attracting growing interest.

The dual valency of double gated devices allows the control of the perpendicular electric field applied onto a few layer graphene. This capability enabled the discovery of an unprecedented way to tune the FLG's low energy band dispersion [59, 63, 108]. To date only the gate tunability of bilayers' band structure has been extensively studied, and a few recent experiments pioneered the study of trilayers electronic properties. However, the gate tunability of thicker few layer graphene with more than three layers has not yet been experimentally studied. This gap in the present knowledge of FLG electronic properties prevents future technological applications from benefitting of the best suited FLG thickness for a given purpose. For instance, although a large gate tunable on/off current ratio in double gated bilayer graphene (up to 100 at room temperature) makes this material system an attractive candidate for transistor applications, it is experimentally unknown whether any thicker few-layer graphene with more than three layers and different stacking orders then Bernal would display an even larger gate-tunable energy band gap as compared to bilayers. It is therefore mandatory to study and compare the gate tunability of FLG electronic properties as a function of the number of layers and of layer stackings as recently theoretically suggested [62, 63, 109].

To date, the gate tunable band gap induced in bilayer graphene makes this material interesting for electronic applications. However, an asymmetry induced by a perpendicular electric field applied onto bilayer graphene, not only opens a gap but affects also the pseudospin of the charge carriers [110]. This pseudospin characterises the layer degree of freedom, and it constitutes an additional quantum number for the charge carriers. Recent theoretical schemes propose the use of the pseudospin for conceptually novel devices in which an on/off state of the current is attained respectively for parallel and antiparallel pseudospin configurations in the bilayer. In these pseudospin-valve devices the polarity of the electric field acting on bilayer graphene plays a similar role as the magnetic field in spin-valve devices. So far, no experimental realization of a pseudospin-valve has been reported, remaining therefore a top priority in exploring the feasibility of the emerging pseudospintronic devices.

In the quest to tune the electronic properties of graphene systems, elemental doping is a recently emerged method alternative to the electrostatic gating (see Fig. 17a). In this case, foreign atoms can substitute carbon atoms or they can be added to the graphene lattice modifying controllably the graphene band structure. Recent experimental studies have already accomplished a band structure modification of graphene by substitution of carbon atoms with boron or nitrogen [111] and by adsorbing atoms on the surface of graphene (see Fig. 17b) [25, 112]. At the same time, the ability to engineer extended defects such as octagonal holes [113] and grain boundaries [114] in graphene crystals is a conceptually novel way to tailor the graphene band structure at the nanoscale. Extended defects (see Fig. 17c) can behave as metallic nanowires and their engineering would allow a variety of novel applications such as guides for charge and spin. Other defects, such as grain boundaries between crystalline graphene domains, are also of particular interest for future device applications on electron lenses since they are predicted to have high transparency or perfect reflection of charge carriers depending on their structure (see Fig. 17d) [114]. Therefore, the ability to engineer dopants and defects in graphene at the atomic level will play a central role in tailoring the electronic structure of graphene systems and will pave a new road towards truly nanometer scale electronic devices.

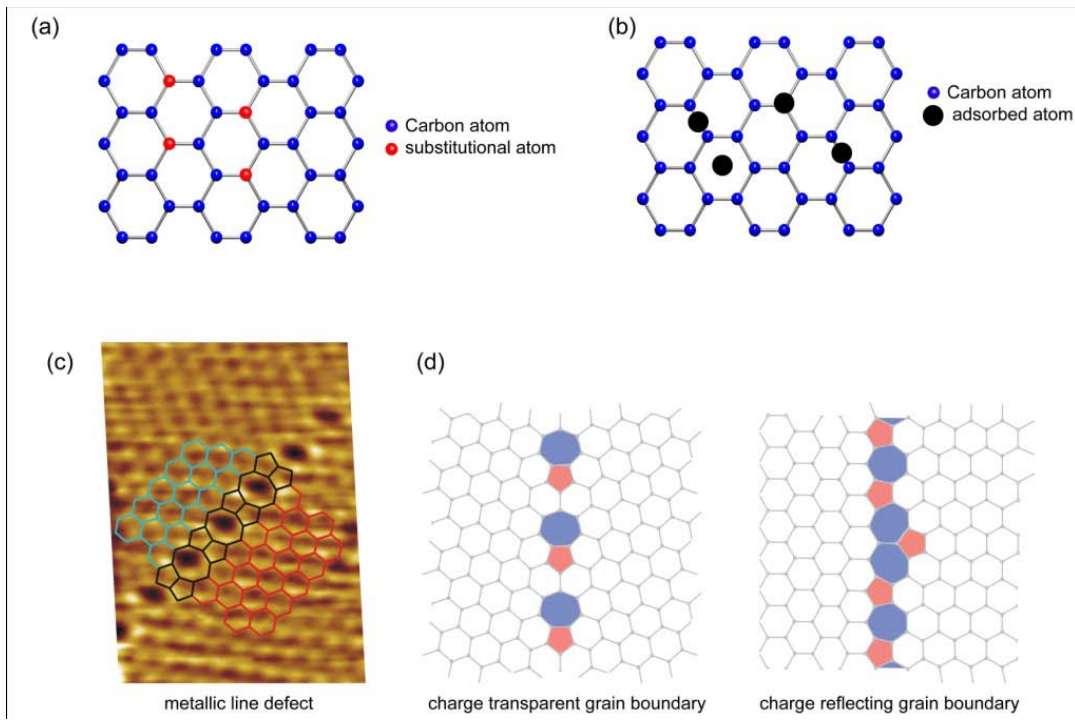

**Figure 17.** Recently emerged methods for tuning the electronic properties of graphene: (a) substitutional doping, (b) adsorbed atoms on graphene, (c) extended defects (From ref. [113]. Copyright 2010 Macmillan Publishers Limited.), and (d) grain boundaries (From ref. [114]. Copyright 2010 Macmillan Publishers Limited.).

**Acknowledgments**
The authors acknowledge E. Mariani, A. Shytov and D.Horsell for careful reading the manuscript and A. A. Kozikov for the measurements of figure 2.

**References**
[1] K.S. Novoselov, A.K. Geim, S.V. Morozov, D. Jiang, Y. Zhang, S.V. Dubonos, et al., Science 306 (2004) 666.
[2] K.S. Novoselov, D. Jiang, T. Booth, V.V. Khotkevich, S.M. Morozov, A.K. Geim, Proc. Natl. Acad. Sci. 102 (2005b) 10451.
[3] K.S. Novoselov, A.K. Geim, S.V. Morozov, D. Jiang, M.I. Katsnelson, I.V. Grigorieva, et al., Nature 438 (2005) 197.
[4] Y.B. Zhang, Y.W. Tan, H.L. Stormer, P. Kim, Nature 438 (2005) 201.
[5] A.K. Geim, K.S. Novoselov, Nature Mater. 6 (2007) 183.
[6] A.H. Castro Neto, F. Guinea, N.M.R. Peres, K.S. Novoselov, A.K. Geim, Rev. Mod. Phys. 81 (2009) 109.
[7] A.K. Geim, Science 324 (2009) 1530.

[8] R.R. Nair, P. Blake, A.N. Grigorenko, K.S. Novoselov, T.J. Booth, T. Stauber, et al., Science 320 (2008) 1308.
[9] K.S. Kim, Y. Zhao, H. Jang, S.Y. Lee, J.M. Kim, K.S. Kim, et al., Nature 457 (2009) 706.
[10] S. Bae, H. Kim, Y. Lee, X. Xu, J.S. Park, Y. Zheng, et al., Nature Nanotech. 5 (2010) 574.
[11] C. Lee, X. Wei, J.W. Kysar, J. Hone, Science 321 (2008) 385.
[12] A.H. Castro Neto, Materials Today 13 (2010) 12.
[13] A.H. Castro Neto, F. Guinea, N.M.R. Peres, Physics Word 11 (2006) 33.
[14] M.S. Fuhrer, G.N. Lau, A.H. MacDonald, MRS Bulletin 35 (2010) 289.
[15] A.K. Geim, A.H. MacDonald, Physics Today 60 (2007) 35.
[16] M.I. Katsnelson, Materials Today 10 (2007) 20.
[17] N.M.R. Peres, Europhysics News 40 (2009) 17.
[18] P.R. Wallace, Phys. Rev. 71 (1947) 622.
[19] M.I. Katsnelson, K.S. Novoselov, A.K. Geim, Nature Phys. 2 (2006) 620; V.V. Cheianov, V.I. Fal'ko, Phys. Rev. B 74 (2006) 041403R.
[20] C. W. J. Beenakker, Rev. Mod. Phys. 80 (2008) 1337.
[21] R.V. Gorbachev, A.S. Mayorov, A.K. Savchenko, D.W. Horsell, F. Guinea, Nano Lett. 8 (2008) 1995.
[22] N. Stander, B. Huard, D. Goldhaber-Gordon, Phys. Rev. Lett. 102 (2009) 026807.
[23] A.F. Young, P. Kim, Nature Phys. 5 (2009) 222.
[24] V.V. Cheianov, V. Fal'ko, B.L. Altshuler, Science 315 (2007) 1252.
[25] T. Ohta, A. Bostowick, T. Seyller, K. Horn, E. Rotenberg, Science 313 (2006) 951.
[26] E.V. Castro, K.S. Novoselov, S.V. Morozov, N.M.R. Peres, J.M.B. Lopes dos Santos, J. Nilsson, et al., Phys. Rev. Lett. 99 (2007) 216802.
[27] J.B. Oostinga, H.B. Heersche, X. Liu, A.F. Morpurgo, L.M.K. Vandersypen, Nature Mater. 7 (2008) 151.
[28] L.M. Zhang, Z.Q. Li, D.N. Basov, M.M. Fogler, Z. Hao, M.C. Martin, Phys. Rev. B 78 (2008) 235408.
[29] S.Y. Zhou, D.A. Siegel, A.V. Fedorov, A. Lanzara, Phys. Rev. Lett. 101 (2008) 086402.
[30] Y.B. Zhang, T.T. Tang, C. Girit, Z. Hao, M.C. Martin, A. Zettl et al., Nature 459 (2009) 820.
[31] K.F. Mak, C.H. Lui, J. Shan, T.F. Heinz, Phys. Rev. Lett. 102 (2009) 256405.
[32] A.B. Kuzmenko, I. Crassee, D. van der Marel, P. Blake, K.S. Novoselov, Phys. Rev. B 80 (2009) 165406.
[33] S. Russo, M.F. Craciun, M. Yamamoto, S. Tarucha, A.F. Morpurgo, New J. Phys. 11 (2009) 095018.
[34] F. Xia, D.B. Farmer, Y. Lin, P. Avouris, Nano Lett. 10 (2010) 715.
[35] K. Zou, J. Zhu, arXiv:1008.0783v1.
[36] M.F. Craciun, S. Russo, M. Yamamoto, J.B. Oostinga, A.F. Morpurgo, S. Tarucha, Nature Nanotechnol. 4 (2009) 383.
[37] J.R. Williams, L. DiCarlo, C.M. Marcus, Science 317 (2007) 638.
[38] B. Huard, J.A. Sulpizio, N. Stander, K. Todd, B. Yang, D. Goldhaber-Gordon, Phys Rev. Lett. 98 (2007) 236803.
[39] B. Ozyilmaz, P. Jarillo-Herrero, D. Efetov, D.A. Abanin, L.S. Levitov, P. Kim, Phys. Rev. Lett. 99 (2007) 166804.
[40] G. Liu, J. Velasco Jr, W. Bao, C.N. Lau, Appl. Phys. Lett. 92 (2008) 203103.
[41] J. Velasco Jr, G. Liu, W. Bao, C.N. Lau, New J. Phys. 11 (2009) 095008.
[42] J. Velasco Jr, G. Liu, L. Jing, P. Kratz, H. Zhang, W. Bao, et al., Phys Rev. B 81 (2010) 121407R.

[43] G.W. Semenoff, Phys. Rev. Lett. 53 (1984) 2449.
[44] Z. Jiang, E.A. Henriksen, L.C. Tung, Y.-J. Wang, M.E. Schwartz, M. Y. Han, et al., Phys. Rev. Lett. 98 (2007) 197403.
[45] Y. Zhang, V.W. Brar, F. Wang, C. Girit, Y. Yayon, M. Panlasigui, et al., Nature Phys. 4 (2008) 627.
[46] T. Ando, T. Nakanishi, R. Saito, J. Phys. Soc. Jpn. 67 (1998) 2857.
[47] H. Suzuura, T. Ando, Phys. Rev. Lett. 89 (2002) 266603.
[48] E. McCann, Phys. Rev. B 74 (2006) 161403R.
[49] E. McCann, V. I. Fal'ko, Phys. Rev. Lett. 96 (2006) 086805.
[50] F. Guinea, A.H. Castro Neto, N.M.R. Peres, Phys. Rev. B 73 (2006) 245426.
[51] M. Koshino, T. Ando, Phys. Rev. B 73 (2006) 245403.
[52] M. Nakamura, E.V. Castro, B. Dóra, Phys. Rev. Lett. 103 (2009) 266804.
[53] S. Latil, L. Henrard, Phys. Rev. Lett. 97 (2006) 036803.
[54] B. Partoens, F.M. Peeters, Phys. Rev. B 74 (2006) 075404.
[55] B. Partoens, F.M. Peeters, Phys. Rev. B 75 (2007) 193402.
[56] M. Koshino, T. Ando, Phys. Rev. B 76 (2007) 085425.
[57] M. Koshino, T. Ando, Phys. Rev. B 77 (2008) 115313.
[58] C.L. Lu, C.P. Chang, Y.C. Huang, R.B. Chen, M.L. Lin, Phys. Rev. B 73 (2006) 144427.
[59] M. Koshino, E. McCann, Phys. Rev. B 79 (2009) 125443.
[60] T. Ohta, A. Bostwick, J.L. McChesney, T. Seyller, K. Horn, E. Rotenberg, Phys. Rev. Lett. 98 (2007) 206802.
[61] M. Aoki, H. Amawashi, Solid State Commun. 142 (2007) 123.
[62] A.A. Avetisyan, B. Partoens, F.M. Peeters, Phys. Rev. B 80 (2009) 195401.
[63] M. Koshino, Phys. Rev. B 81 (2010) 125304.
[64] M. Koshino, E. McCann, Phys. Rev. B 81 (2010) 115315.
[65] H. Lipson, A. R. Stokes, Proc. R. Soc. London A181 (1942) 101.
[66] N.M.R. Peres, arXiv:1007.2849v1
[67] M.I. Katsnelson, Eur. Phys. J. B 51 (2006b) 157.
[68] J. Tworzydlo, B. Trauzettel, M. Titov, A. Rycerz, C.W.J. Beenakker, Phys. Rev. Lett. 96 (2006) 246802.
[69] M.H. Johnson, B. A. Lippmann, Phys. Rev. 76 (1949) 828.
[70] M.M. Nieto, P.L. Taylor, Am. J. Phys. 53 (1985) 234.
[71] V.P. Gusynin, S.G. Sharapov, Phys. Rev. Lett. 95 (2005) 146801.
[72] N.M.R. Peres, F. Guinea, A.H. Castro Neto, Phys. Rev. B 73 (2006) 125411.
[73] K.S. Novoselov, E. McCann, S.V. Morozov, V.I. Fal'ko, M.I. Katsnelson, U. Zeitler, et al., Nature Phys. 2 (2006) 177.
[74] B.N. Szafranek, D. Schall, M. Otto, D. Neumaier, H. Kurz, App. Phys. Lett. 96 (2010) 112103.
[75] M.C. Lemme, T.J. Echtermeyer, M. Baus, B.N. Szafranek, J. Bolten, M. Schmidt, et al., Solid State Electronics 52 (2008) 514.
[76] T.J. Echtermeyer, M.C. Lemme, J. Bolten, M. Baus, M. Ramsteiner, M. Ramsteiner, et al., European Physical Journal-Special Topics 148 (2007) 19.
[77] X. Li, W. Cai, J. An, S. Kim, J. Nah, D. Yang, et al., Science 324 (2009) 1312.
[78] A.N. Pal, A. Ghosh Phys. Rev. Lett. 102 (2009) 126805.
[79] H. Miyazaki, S. Odaka, T. Sato, S. Tanaka, H. Goto, A. Kanda, et al., Applied Physics Express 1 (2008) 034007.
[80] S. Kim, J. Nah, I. Jo, D, Shahrjerdi, L. Colombo, Z. Yao et al., Appl. Phys. Lett. 94 (2009) 062107.
[81] D.B. Farmer, Y.-M. Lin, P. Avouris, Appl. Phys. Lett. 97 (2010) 013103.

[82] Z. Wang, H. Xu, Z. Zhang, S.Wang, L. Ding, Q. Zeng, et al., Nano Lett. 10 (2010) 2024.
[83] D. Kondo, S. Sato, K. Yagi, N. Harada, M. Sato, M. Nihei, et al., Applied Physics Express 3 (2010) 025102.
[84] D.B. Farmer et al., Nano Lett. 9, 4474 (2009).
[85] I. Meric, M.Y. Han, A.F. Young, B. Ozyilmas, P. Kim, K.L. Shepard, Nature Nanotech. 3 (2008) 654.
[86] B. Ozyilmaz, P. Jarillo-Herrero, D. Efetov, P. Kim, Appl. Phys. Lett. 91 (2007) 192107.
[87] L. Liao, J.W. Bai, R. Cheng, Y.C. Lin, S. Jiang, Y. Huang, et al., Nano Lett. 10 (2010) 1917.
[88] F. Schwierz, Nature Nanotech. 5 (2010) 487.
[89] Y. Ohno, K. Maehashi, Y. Yamashiro, K. Matsumoto, Nano Lett. 9 (2009) 3318.
[90] N.B. Kopnin, E. B. Sonin, Phys. Rev. Lett. 100 (2008) 246808.
[91] B. Uchoa, A.H. Castro Neto, Phys. Rev. Lett. 98 (2007) 146801.
[92] H. Yuan, H. Shimotani, A. Tsukazaki, A. Ohtomo, M. Kawasaki, Y. Iwasa, Adv. Funct. Mater. 19 (2009) 1046.
[93] R.W.I de Boer, A.F.Stassen, M.F. Craciun, C.L. Mulder, A. Molinari, S. Rogge, et al., App. Phys. Lett. 86 (2005) 262109.
[94] J. Zamuseil, H. Sirringhaus, Chem. Rev. 107 (2007) 1296.
[95] D.A. Abanin, L. S. Levitov, Science 317 (2007) 641.
[96] O. Klein, Z. Phys. 53 (1929) 157.
[97] A.V. Shytov, M.S. Rudner, L.S. Levitov, Phys. Rev. Lett. 101 (2008) 156804.
[98] V.M. Pereira, A.H. Castro Neto, Phys. Rev Lett. 103 (2009) 046801.
[99] F. Guinea, M.I Katsnelson, A.K. Geim, Nature Physics 6 (2009) 30.
[100] G Giovannetti, P.A. Khomyakov, G. Brocks, P.J. Kelly, J. van den Brink, Phys. Rev. B 76 (2007) 073103.
[101] L. Ci, L. Song, C.H. Jin, D. Jariwala, D.X. Wu, Y.J. Li, et al., Nature Mater. 9 (2010) 430.
[102] Y.W. Son, M.L. Cohen, S.G. Louie, Phys. Rev Lett. 97 (2006) 216803.
[103] D.C. Elias, R.R. Nair, T.M.G. Mohiuddin, S.V. Morozov, P. Blake, M.P. Halsall, et al., Science 323 (2009) 610; R. Balog, B. Jorgensen, L. Nilsson, M. Andersen, E. Rienks, M. Bianchi, et al., Nature Mater. 9 (2010) 315; F. Withers, M. Dubois, A.K.Savchenko, Phys. Rev. B 82 (2010) 073403; R.R. Nair, W.C. Ren, R. Jalil, I. Riaz, V.G. Kravets, L. Britnell, et al., arxiv:1006.3016.
[104] E.V. Castro, K.S. Novoselov, S.V. Morozov, N.M.R Peres, J.M.B. Lopes dos Santos, J. Nilsson, et al., J. Phys.: Condens. Matter 22 (2010) 175503.
[105] V.V. Mkhitaryan, M.E. Raikh, Phys. Rev. B 78 (2008) 195409; M. Koshino, Phys. Rev. B 78 (2008) 155411.
[106] K. Nagashio, T. Nishimura, K. Kita, A. Toriumi, Appl. Phys. Express 2 (2009) 025003.
[107] J. Martin, N. Akerman, G. Ulbricht, T. Lohmann, J.H. Smet, K. von Klitzing, et al., Nature Phys. 4 (2008) 144.
[108] A.A. Avetisyan, B. Partoens, F.M. Peeters, Phys. Rev. B 80 (2009) 195401.
[109] M. Koshino, E. McCann, Phys. Rev. B 80 (2009) 165409.
[110] H. Min, G. Borghi, M. Polini, A.H. MacDonald, Phys. Rev. B 77 (2008) 041407R; P. San-Jose, E. Prada, E. McCann, H. Schomerus, Phys. Rev. Lett. 102 (2009) 247204.
[111] X. Wang, X. Li, L. Zhang, Y. Yoon, P. K. Weber, H. Wang, et al., Science 324 (2009) 768; D. Wei, Y. Liu, Y. Wang, H. Zhang, L. Huang, G. Yu, Nano Lett. 9 (2009) 1752; L. S. Panchakarla, K. S. Subrahmanyam, S. K. Saha, A. Govindaraj, H. R. Krishnamurthy, U. V. Waghmare, et al., Adv. Mater. 21 (2009) 4726.

[112] I. Gierz, C. Riedl, U. Starke, C. R. Ast, K. Kern, Nano Lett. 8 (2008) 4603; Y. Shi, K. K. Kim, A. Reina, M. Hofmann, L.-J. Li, J. Kong, Nano Lett. 4 (2010) 2689.
[113] J. Lahiri, Y. Lin, P. Bozkurt, I. I. Oleynik, M. Batzill, Nature Nanotech. 5 (2010) 326.
[114] O. V. Yazyev, S. G. Louie, Nature Mater. 9 (2010) 806.